\pdfoutput=1

\documentclass[12pt,a4paper]{article}

\usepackage{float} 
\usepackage{ifthen} 
\newboolean{pdflatex}
\setboolean{pdflatex}{true} 

\newboolean{articletitles}
\setboolean{articletitles}{true} 

\newboolean{uprightparticles}
\setboolean{uprightparticles}{false} 

\newboolean{inbibliography}
\setboolean{inbibliography}{true} 

\usepackage{microtype}
\usepackage{lineno}  
\usepackage{xspace} 

\usepackage{graphicx}  
\usepackage{color}
\usepackage{colortbl}
\graphicspath{{./figs/}} 

\usepackage{amsmath} 
\usepackage{amssymb}
\usepackage{amsfonts}
\usepackage{upgreek} 

\newcommand*\patchAmsMathEnvironmentForLineno[1]{%
\expandafter\let\csname old#1\expandafter\endcsname\csname #1\endcsname
\expandafter\let\csname oldend#1\expandafter\endcsname\csname
end#1\endcsname
 \renewenvironment{#1}%
   {\linenomath\csname old#1\endcsname}%
   {\csname oldend#1\endcsname\endlinenomath}%
}
\newcommand*\patchBothAmsMathEnvironmentsForLineno[1]{%
  \patchAmsMathEnvironmentForLineno{#1}%
  \patchAmsMathEnvironmentForLineno{#1*}%
}
\AtBeginDocument{%
\patchBothAmsMathEnvironmentsForLineno{equation}%
\patchBothAmsMathEnvironmentsForLineno{align}%
\patchBothAmsMathEnvironmentsForLineno{flalign}%
\patchBothAmsMathEnvironmentsForLineno{alignat}%
\patchBothAmsMathEnvironmentsForLineno{gather}%
\patchBothAmsMathEnvironmentsForLineno{multline}%
}

\usepackage{hyperref}    
\usepackage[all]{hypcap} 

\def\lhcb {\mbox{LHCb}\xspace}

\ifthenelse{\boolean{uprightparticles}}%
{

 \def\Ppi         {\ensuremath{\uppi}\xspace}

 \def\PDelta      {\ensuremath{\Delta}\xspace}                 
 \def\PXi      {\ensuremath{\Xi}\xspace}                 
 \def\PLambda      {\ensuremath{\Lambda}\xspace}                 
 \def\PSigma      {\ensuremath{\Sigma}\xspace}                 
 \def\POmega      {\ensuremath{\Omega}\xspace}                 
 \def\PUpsilon      {\ensuremath{\Upsilon}\xspace}                 
 
 \def\PB      {\ensuremath{\mathrm{B}}\xspace}                 
                  
 \def\PD      {\ensuremath{\mathrm{D}}\xspace}

 \def\PK      {\ensuremath{\mathrm{K}}\xspace}

 \def\Pb      {\ensuremath{\mathrm{b}}\xspace}                 
 \def\Pc      {\ensuremath{\mathrm{c}}\xspace}

 \def\Pi      {\ensuremath{\mathrm{i}}\xspace}

 \def\Ps      {\ensuremath{\mathrm{s}}\xspace}

}
{

 \def\Ppi         {\ensuremath{\pi}\xspace}

 \mathchardef\PDelta="7101
 \mathchardef\PXi="7104
 \mathchardef\PLambda="7103
 \mathchardef\PSigma="7106
 \mathchardef\POmega="710A
 \mathchardef\PUpsilon="7107
                  
 \def\PB      {\ensuremath{B}\xspace}                 
                  
 \def\PD      {\ensuremath{D}\xspace}

 \def\PK      {\ensuremath{K}\xspace}

 \def\Pb      {\ensuremath{b}\xspace}                 
 \def\Pc      {\ensuremath{c}\xspace}

 \def\Pi      {\ensuremath{i}\xspace}

 \def\Ps      {\ensuremath{s}\xspace}

}

\def\squark    {\ensuremath{\Ps}\xspace}

\def\cquark    {\ensuremath{\Pc}\xspace}

\def\bquark    {\ensuremath{\Pb}\xspace}

\def\pion  {\ensuremath{\Ppi}\xspace}

\def\pip   {\ensuremath{\pion^+}\xspace}
\def\pim   {\ensuremath{\pion^-}\xspace}

\def\kaon  {\ensuremath{\PK}\xspace}
  \def\Kbar  {\kern 0.2em\overline{\kern -0.2em \PK}{}\xspace}

\def\Km    {\ensuremath{\kaon^-}\xspace}

  \def\Dbar    {\kern 0.2em\overline{\kern -0.2em \PD}{}\xspace}
\def\D       {\ensuremath{\PD}\xspace}

\def\Dz      {\ensuremath{\D^0}\xspace}

\def\Dp      {\ensuremath{\D^+}\xspace}
\def\Dm      {\ensuremath{\D^-}\xspace}

\def\Ds      {\ensuremath{\D^+_\squark}\xspace}
\def\Dsp     {\ensuremath{\D^+_\squark}\xspace}

\def\Bbar    {\ensuremath{\kern 0.18em\overline{\kern -0.18em \PB}{}}\xspace}

  \def\Y#1S{\ensuremath{\PUpsilon{(#1S)}}\xspace}

\def\Lbar {\ensuremath{\kern 0.1em\overline{\kern -0.1em\PLambda}}\xspace}

\newcommand{\decay}[2]{\ensuremath{#1\!\to #2}\xspace}         

\def\to                 {\ensuremath{\rightarrow}\xspace}

\def\CP                {\ensuremath{C\!P}\xspace}

\def\CPV               {\ensuremath{C\!PV}\xspace}

\def\SCPi              {\ensuremath{\mathcal{S}^i_{C\!P}}\xspace}

\def\Dppp         {\decay{\Dp}{\pim\pip\pip}}
\def\DKpp         {\decay{\Dp}{\Km\pip\pip}}
\def\Dsppp        {\decay{\Ds}{\pim\pip\pip}}

\def\AT#1     {\ensuremath{A_{\mathrm{T}}^{#1}}\xspace}           

\def\C#1      {\ensuremath{\mathcal{C}_{#1}}\xspace}                       
\def\Cp#1     {\ensuremath{\mathcal{C}_{#1}^{'}}\xspace}                    
\def\Ceff#1   {\ensuremath{\mathcal{C}_{#1}^{\mathrm{(eff)}}}\xspace}        
\def\Cpeff#1  {\ensuremath{\mathcal{C}_{#1}^{'\mathrm{(eff)}}}\xspace}       
\def\Ope#1    {\ensuremath{\mathcal{O}_{#1}}\xspace}                       
\def\Opep#1   {\ensuremath{\mathcal{O}_{#1}^{'}}\xspace}                    

\newcommand{\tev}{\ifthenelse{\boolean{inbibliography}}{\ensuremath{~T\kern -0.05em eV}\xspace}{\ensuremath{\mathrm{\,Te\kern -0.1em V}}\xspace}}
\newcommand{\gev}{\ensuremath{\mathrm{\,Ge\kern -0.1em V}}\xspace}
\newcommand{\mev}{\ensuremath{\mathrm{\,Me\kern -0.1em V}}\xspace}
\newcommand{\kev}{\ensuremath{\mathrm{\,ke\kern -0.1em V}}\xspace}
\newcommand{\ev}{\ensuremath{\mathrm{\,e\kern -0.1em V}}\xspace}
\newcommand{\gevc}{\ensuremath{{\mathrm{\,Ge\kern -0.1em V\!/}c}}\xspace}
\newcommand{\mevc}{\ensuremath{{\mathrm{\,Me\kern -0.1em V\!/}c}}\xspace}
\newcommand{\gevcc}{\ensuremath{{\mathrm{\,Ge\kern -0.1em V\!/}c^2}}\xspace}
\newcommand{\gevgevcccc}{\ensuremath{{\mathrm{\,Ge\kern -0.1em V^2\!/}c^4}}\xspace}
\newcommand{\mevcc}{\ensuremath{{\mathrm{\,Me\kern -0.1em V\!/}c^2}}\xspace}

\def\mum  {\ensuremath{\,\upmu\rm m}\xspace}

\def\gsim{{~\raise.15em\hbox{$>$}\kern-.85em
          \lower.35em\hbox{$\sim$}~}\xspace}
\def\lsim{{~\raise.15em\hbox{$<$}\kern-.85em
          \lower.35em\hbox{$\sim$}~}\xspace}

\def\pt         {\mbox{$p_{\rm T}$}\xspace}

\def\tell1  {TELL1\xspace}
\def\ukl1   {UKL1\xspace}

\usepackage{cite} 
\usepackage{mciteplus}

\usepackage{longtable} 

\begin{document}

\renewcommand{\thefootnote}{\fnsymbol{footnote}}
\setcounter{footnote}{1}


\begin{titlepage}
\pagenumbering{roman}

\vspace*{-1.5cm}
\centerline{\large EUROPEAN ORGANIZATION FOR NUCLEAR RESEARCH (CERN)}
\vspace*{1.5cm}
\hspace*{-0.5cm}
\begin{tabular*}{\linewidth}{lc@{\extracolsep{\fill}}r}
\ifthenelse{\boolean{pdflatex}}
{\vspace*{-2.7cm}\mbox{\!\!\!\includegraphics[width=.14\textwidth]{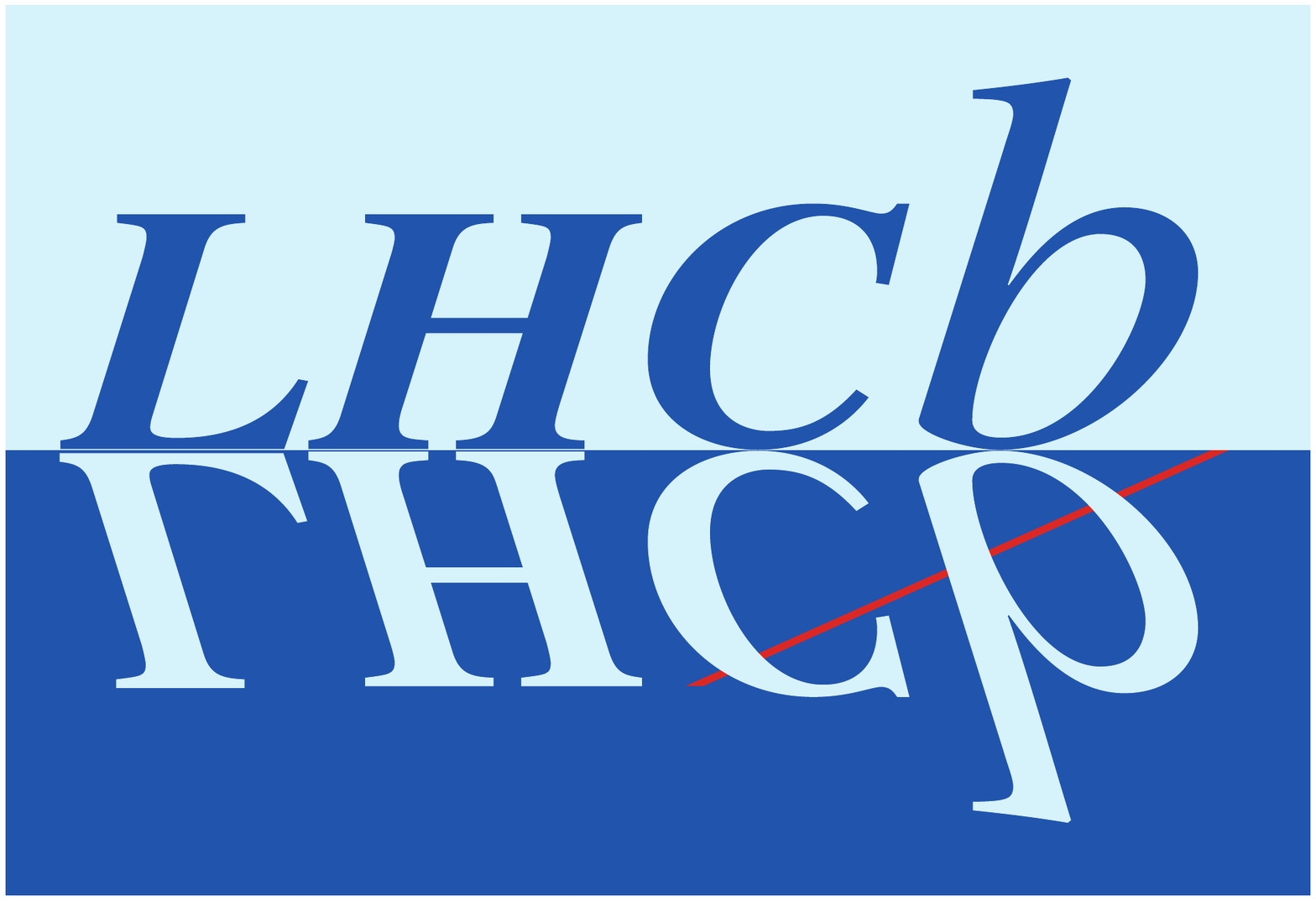}} & &}%
{\vspace*{-1.2cm}\mbox{\!\!\!\includegraphics[width=.12\textwidth]{figs/lhcb-logo.eps}} & &}%
\\
 & & CERN-PH-EP-2013-199 \\  
 & & LHCb-PAPER-2013-057 \\  
 & & 29 October 2013 \\ 
\end{tabular*}

\vspace*{4.0cm}

{\bf\boldmath\huge
\begin{center}
  Search for \CP violation in the decay $D^+ \to \pi^-\pi^+\pi^+$
\end{center}
}

\vspace*{2.0cm}

\begin{center}
The LHCb collaboration\footnote{Authors are listed on the following pages.}
\end{center}

\vspace{\fill}

\begin{abstract}
  \noindent
  A search for \CP violation in the phase space of the decay $D^+\to\pi^-\pi^+\pi^+$ is reported using $pp$ collision data, corresponding to an integrated luminosity of 
  1.0~fb$^{-1}$, collected by the LHCb experiment at  a centre-of-mass energy of 7 TeV.    The Dalitz plot distributions for   $3.1\times 10^6$  $D^+$ and $D^-$ candidates are compared with binned and unbinned model-independent techniques.  No evidence for \CP violation is found.
\end{abstract}

\vspace*{2.0cm}

\begin{center}
  Submitted to Phys.~Lett.~B 
  \end{center}

\vspace{\fill}

{\footnotesize 
\centerline{\copyright~CERN on behalf of the \lhcb collaboration, license \href{http://creativecommons.org/licenses/by/3.0/}{CC-BY-3.0}.}}
\vspace*{2mm}

\end{titlepage}


\newpage
\setcounter{page}{2}
\mbox{~}
\newpage

\centerline{\large\bf LHCb collaboration}
\begin{flushleft}
\small
R.~Aaij$^{40}$, 
B.~Adeva$^{36}$, 
M.~Adinolfi$^{45}$, 
C.~Adrover$^{6}$, 
A.~Affolder$^{51}$, 
Z.~Ajaltouni$^{5}$, 
J.~Albrecht$^{9}$, 
F.~Alessio$^{37}$, 
M.~Alexander$^{50}$, 
S.~Ali$^{40}$
% ...

{\footnotesize \it
$ ^{1}$Centro Brasileiro de Pesquisas F\'{i}sicas (CBPF), Rio de Janeiro, Brazil\\
% ...
$ ^{5}$Clermont Universit\'{e}, Universit\'{e} Blaise Pascal, CNRS/IN2P3, LPC, Clermont-Ferrand, France\\
$ ^{6}$CPPM, Aix-Marseille Universit\'{e}, CNRS/IN2P3, Marseille, France\\
% ...
$ ^{9}$Fakult\"{a}t Physik, Technische Universit\"{a}t Dortmund, Dortmund, Germany\\
% ...
$ ^{36}$Universidad de Santiago de Compostela, Santiago de Compostela, Spain\\
$ ^{37}$European Organization for Nuclear Research (CERN), Geneva, Switzerland\\
% ...
$ ^{40}$Nikhef National Institute for Subatomic Physics, Amsterdam, The Netherlands\\
% ...
$ ^{45}$H.H. Wills Physics Laboratory, University of Bristol, Bristol, United Kingdom\\
% ...
$ ^{50}$School of Physics and Astronomy, University of Glasgow, Glasgow, United Kingdom\\
$ ^{51}$Oliver Lodge Laboratory, University of Liverpool, Liverpool, United Kingdom\\
% ...
}
\end{flushleft}

\cleardoublepage


\renewcommand{\thefootnote}{\arabic{footnote}}
\setcounter{footnote}{0}



\pagestyle{plain} 
\setcounter{page}{1}
\pagenumbering{arabic}


%


\section{Introduction}
\label{sec:Introduction}

In the  Standard Model (SM) charge-parity (\CP) violation in the charm sector 
 is expected to be  small.
Quantitative predictions of \CP asymmetries
are difficult, since  the computation of strong-interaction effects
in the non-perturbative regime is involved. In spite of this, it
was commonly assumed that the observation   of asymmetries of the order
of 1\% in charm decays would be an indication of new sources of \CP violation (\CPV).
Recent studies, however, suggest that \CP asymmetries of this magnitude
could still be accommodated within the SM \cite{grossman,brod,gronau,LHCb-PAPER-2012-031}.

Experimentally, the sensitivity for \CPV searches has substantially increased 
 over the past few years. Especially with the advent of 
 the large LHCb data set,  \CP asymmetries at the $\mathcal{O}(10^{-2})$ 
level are disfavoured \cite{LHCb-PAPER-2012-052,LHCb-CONF-2013-003, CDF-PRLDACP, LHCb-PAPER-2013-053, LHCb-PAPER-2013-054}. 
 With uncertainties approaching 
$\mathcal{O}(10^{-3})$, the current \CPV searches  start to probe the regime 
of the SM expectations.

The most simple and direct technique for \CPV searches is 
the computation of an asymmetry between the particle and anti-particle time-integrated decay rates. 
A single number, however,  may not be sufficient for a comprehension of the nature 
of the \CP violating asymmetry. In this context, three- and four-body decays benefit from  rich resonance structures with 
interfering amplitudes modulated by strong-phase variations across the phase space. Searches for localised asymmetries can bring complementary information on the nature of the \CPV.

In this Letter, a search for \CP violation in the Cabibbo-suppressed decay 
$D^+ \to \pi^-\pi^+\pi^+$ is reported.\footnote{Unless stated explicitly, the inclusion of 
charge conjugate states is implied.}
The investigation is performed across the Dalitz plot using two model-independent techniques,  a binned search  as employed in previous LHCb analyses~\cite{LHCb-PAPER-2011-017,LHCb-PAPER-2013-041} and an unbinned
search based on the nearest-neighbour method \cite{henze,schilling}. 
Possible localised charge asymmetries arising from production or detector effects are investigated using the decay \decay{\Dsp}{\pim\pip\pip}, which has the same final state particles as the signal mode, as a control channel. 
Since it is a Cabibbo-favoured decay,  with negligible loop (penguin) contributions, \CP violation is not  expected at any significant level.

\section{LHCb detector and data set}
\label{sec:Detector}

The \lhcb detector~\cite{Alves:2008zz} is a single-arm forward
spectrometer covering the \mbox{pseudorapidity} range $2<\eta <5$,
designed for the study of particles containing \bquark or \cquark
quarks. The detector includes a high-precision tracking system
consisting of a silicon-strip vertex detector surrounding the $pp$
interaction region, a large-area silicon-strip detector located
upstream of a dipole magnet with a bending power of about
$4{\rm\,Tm}$, and three stations of silicon-strip detectors and straw
drift tubes placed downstream. 
The combined tracking system provides a momentum measurement with
relative uncertainty that varies from 0.4\% at 5\gevc to 0.6\% at 100\gevc,
and impact parameter (IP) resolution of 20\mum for
tracks with high transverse momentum, \pt. Charged hadrons are identified
using two ring-imaging Cherenkov (RICH) detectors  \cite{LHCb-DP-2012-003}. 
Photon, electron and
hadron candidates are identified by a calorimeter system consisting of
scintillating-pad and preshower detectors, an electromagnetic
calorimeter and a hadronic calorimeter. Muons are identified by a
system composed of alternating layers of iron and multiwire
proportional chambers~\cite{LHCb-DP-2012-002}.
The trigger~\cite{LHCb-DP-2012-004} consists of a
hardware stage, based on information from the calorimeter and muon
systems, followed by a software stage, which applies  full event
reconstruction. At the hardware trigger stage,  events are required to have muons with  high transverse momentum or hadrons, photons or electrons with  high transverse energy deposit in the calorimeters. For hadrons, the transverse energy threshold is 3.5\gevcc.

The software trigger  requires at least 
one  good quality track from the signal decay with 
high \pt and high  $\chi^2_{\rm IP}$,  defined as the difference 
in $\chi^2$ of the primary vertex (PV)   reconstructed with and without this particle. A  secondary vertex  is  formed by
three tracks with good quality, each not pointing to any PV,  and with requirements on \pt, momentum $p$, scalar sum of \pt of the tracks, and 
 a significant displacement from any PV.

The data sample used in this analysis corresponds to an integrated luminosity
of 1.0\,fb$^{-1}$ of $pp$ collisions at a centre-of-mass energy of 7 TeV
collected by the LHCb experiment in 2011. The magnetic field polarity is reversed regularly during the data taking in order to minimise
effects of charged particle and antiparticle detection asymmetries. 
Approximately half of the data are collected with each polarity, hereafter referred to as ``magnet up" and 
``magnet down" data.


\section{Event selection}
\label{sec:selection}

To reduce the combinatorial background,  requirements  on the quality of the reconstructed tracks, their $\chi^2_{\rm IP}$, \pt,  and scalar \pt sum are applied.
Additional requirements are made on the secondary vertex fit quality, the minimum significance of the displacement from  the secondary to any primary vertex in the event, and
 the $\chi^2_{\rm IP}$ of the $D^+_{(s)}$ candidate. This also reduces the contribution of secondary \D mesons from \bquark-hadron decays to 1--2\%, 
 avoiding the introduction of new sources of asymmetries. The final-state particles are required to satisfy particle identification (PID) criteria based on the RICH detectors.

After these requirements, there is still a significant background contribution, which
could introduce charge asymmetries across the Dalitz plot. This includes semileptonic decays like $D^+ \to K^-\pi^+\mu^+ \nu$ and $D^+ \to \pi^-\pi^+\mu^+ \nu$;
three-body decays, such as $\D^+ \to K^- \pi^+ \pi^+$; prompt two-body $D^0$ decays forming a three-prong vertex with a 
random pion; and $D^0$ decays from the $D^{*+}$ chain, such as 
$D^{*+} \to D^0(K^-\pi^+,\pi^-\pi^+,K^-\pi^+\pi^0) \pi^+$. The contribution from \DKpp and prompt \Dz decays that involve the misidentification of the kaon as a pion is reduced
to a negligible level with a more stringent PID requirement on the $\pi^-$ candidate.  
The remaining background from semileptonic decays is controlled by applying a muon
veto to all three tracks, using  information from the muon system~\cite{LHCb-DP-2013-001}.
The contribution from the $D^{*+}$ decay chain is reduced to a negligible level with a requirement  on $\chi^2_{\rm IP}$ of the $\pi^+$ candidate with  lowest \pt. 

Fits to the invariant mass distribution $M({\pi^-\pi^+\pi^+})$ are performed for the \Dp and \Ds candidates satisfying the above selection criteria and within the range 
$1810 <M({\pi^-\pi^+\pi^+})< 1930\mevcc$ and $1910<M({\pi^-\pi^+\pi^+})< 2030\mevcc$, respectively. The signal   is described by a sum of two Gaussian functions and the background is represented by a third-order polynomial. The data sample is separated according to magnet polarity and  candidate momentum ($p_{\Dp_{(s)}}\!\!<\!50$\gevc,  $50\!<p_{\Dp_{(s)}}\!\!<\!100$\gevc,  and $p_{\Dp_{(s)}}\!\!>\!100$\gevc), to take into account the dependence of the mass resolution on the momentum. The  parameters are determined  by simultaneous fits to these $D^+_{(s)}$ and $D^-_{(s)}$ subsamples.

\begin{figure}[H]
\begin{center}
\includegraphics*[width=0.48\textwidth]{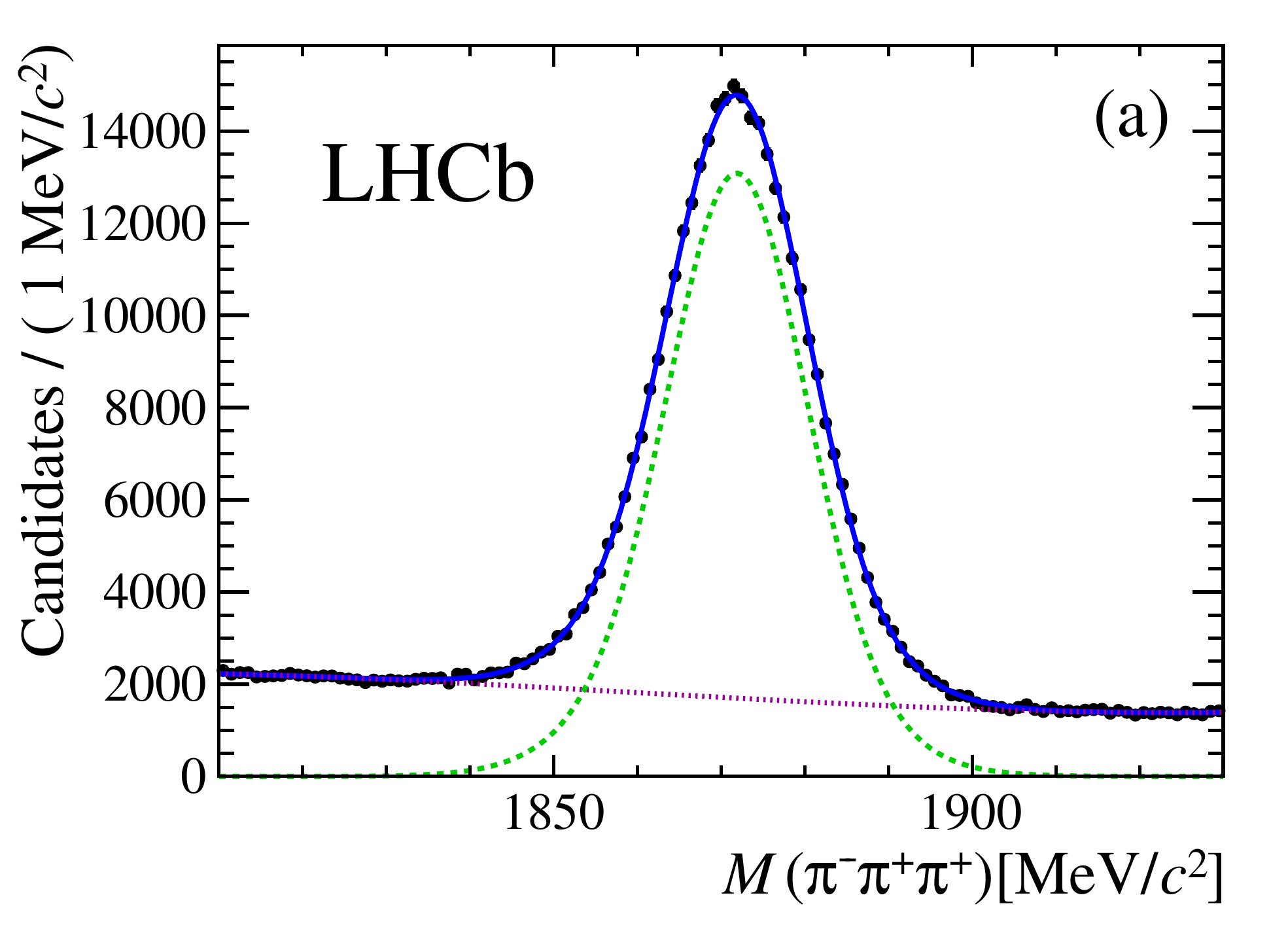}~~
\includegraphics*[width=0.48\textwidth]{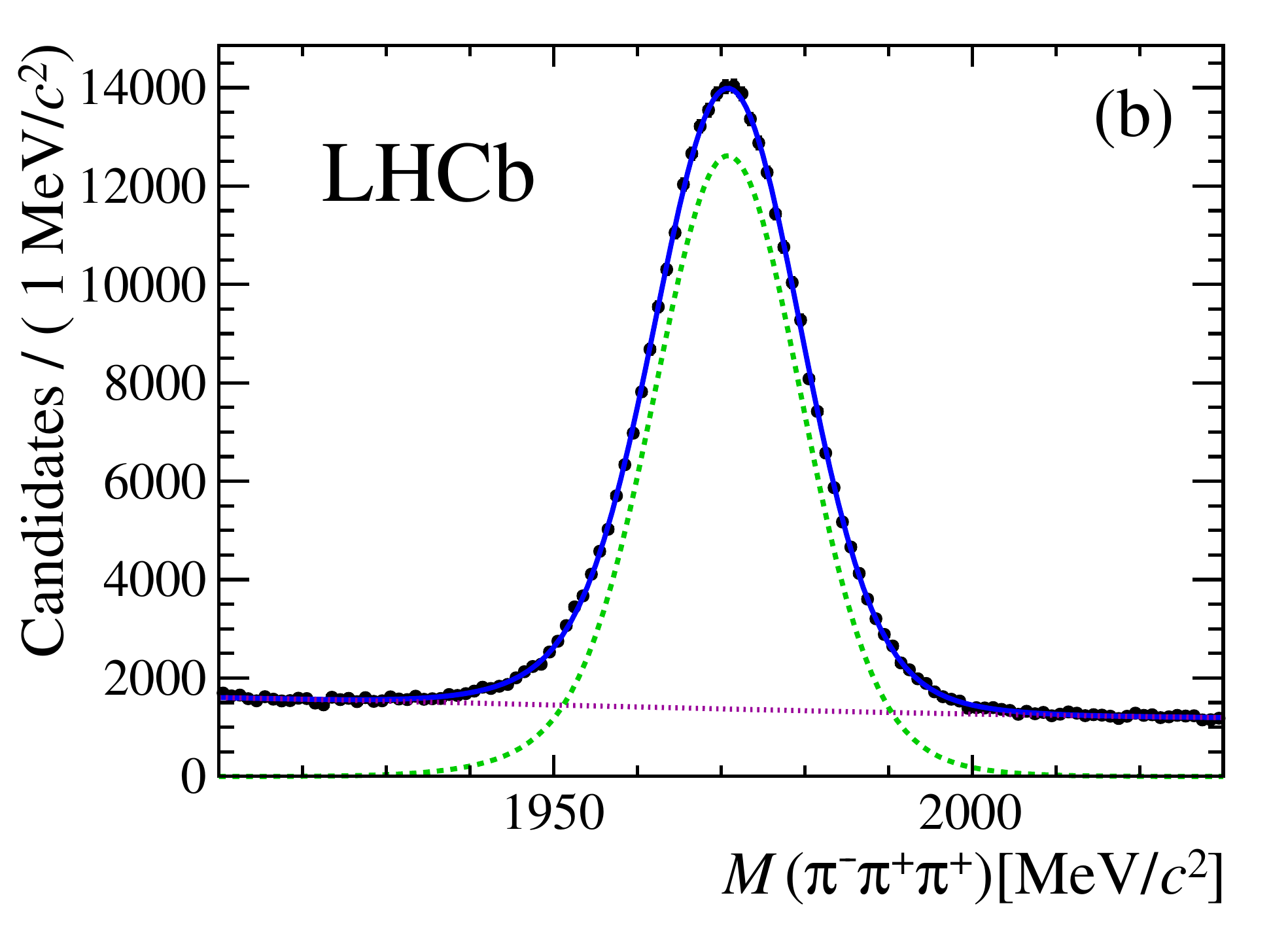} \\
\vspace*{-0.3cm}
\caption{\small Invariant-mass distributions for (a) \Dp and (b) \Ds candidates in the momentum range $50<p_{\Dp_{(s)}}\!\!<100$ \gevc for magnet up data. Data points are shown in black. The solid (blue) line is the  fit function, the (green) dashed line is the signal component   and the (magenta) dotted line is the background.}
\end{center}
\label{fig:fits}
\end{figure}

The \Dp and \Ds invariant mass distributions and fit results for the momentum range $50\! <\!p_{\Dp_{(s)}}\!\!<\!100$\gevc are shown in Fig.~\ref{fig:fits} for magnet up data. The total yields after summing over all fits are 
$(2678\pm 7)\!\! \times\! \! 10^3$ \Dppp and $(2704\pm 8)\!\!\times\!\! 10^3$ \Dsppp decays.  
The final samples  used for the \CPV search consist of all candidates  with $M({\pi^-\pi^+\pi^+})$  within $\pm2\tilde\sigma$ around  $\tilde m_{D_{(s)}}$, where $\tilde\sigma$ and $\tilde m_{D_{(s)}}$ are the weighted average of the two fitted Gaussian widths and mean values. The values of $\tilde\sigma$ range from 8 to 12 \mevcc, depending  on the 
momentum region.  
For the signal sample there are $3114\!\times \!10^3$ candidates, including background, while for the control mode there are $2938 \!\times \!10^3$ candidates with  purities of 82\% and 87\%, respectively. The purity is defined as the fraction of signal decays in this mass range.

\begin{figure}[H]
\begin{center}
\includegraphics*[width=0.45\textwidth]{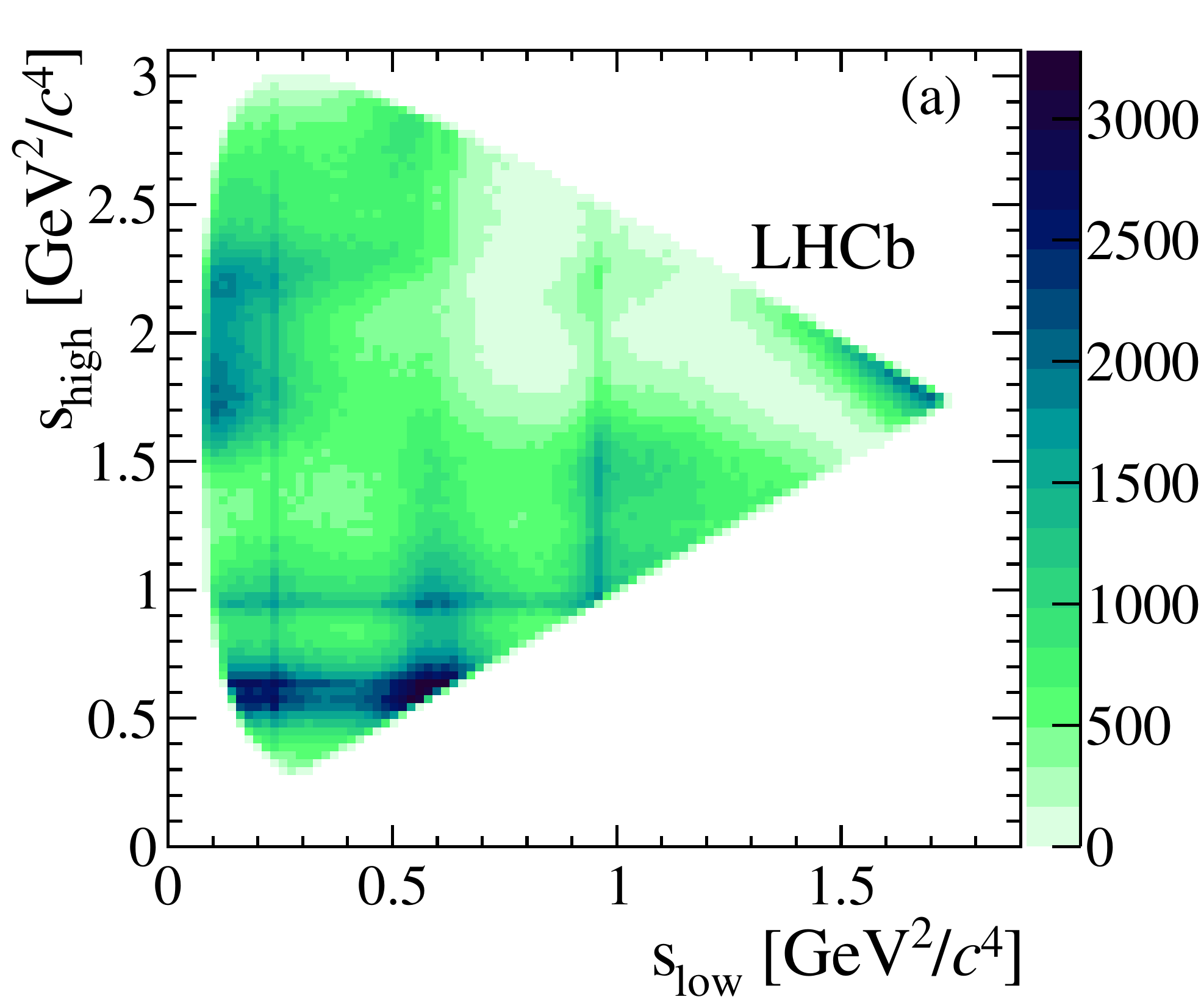}~~~
\includegraphics*[width=0.45\textwidth]{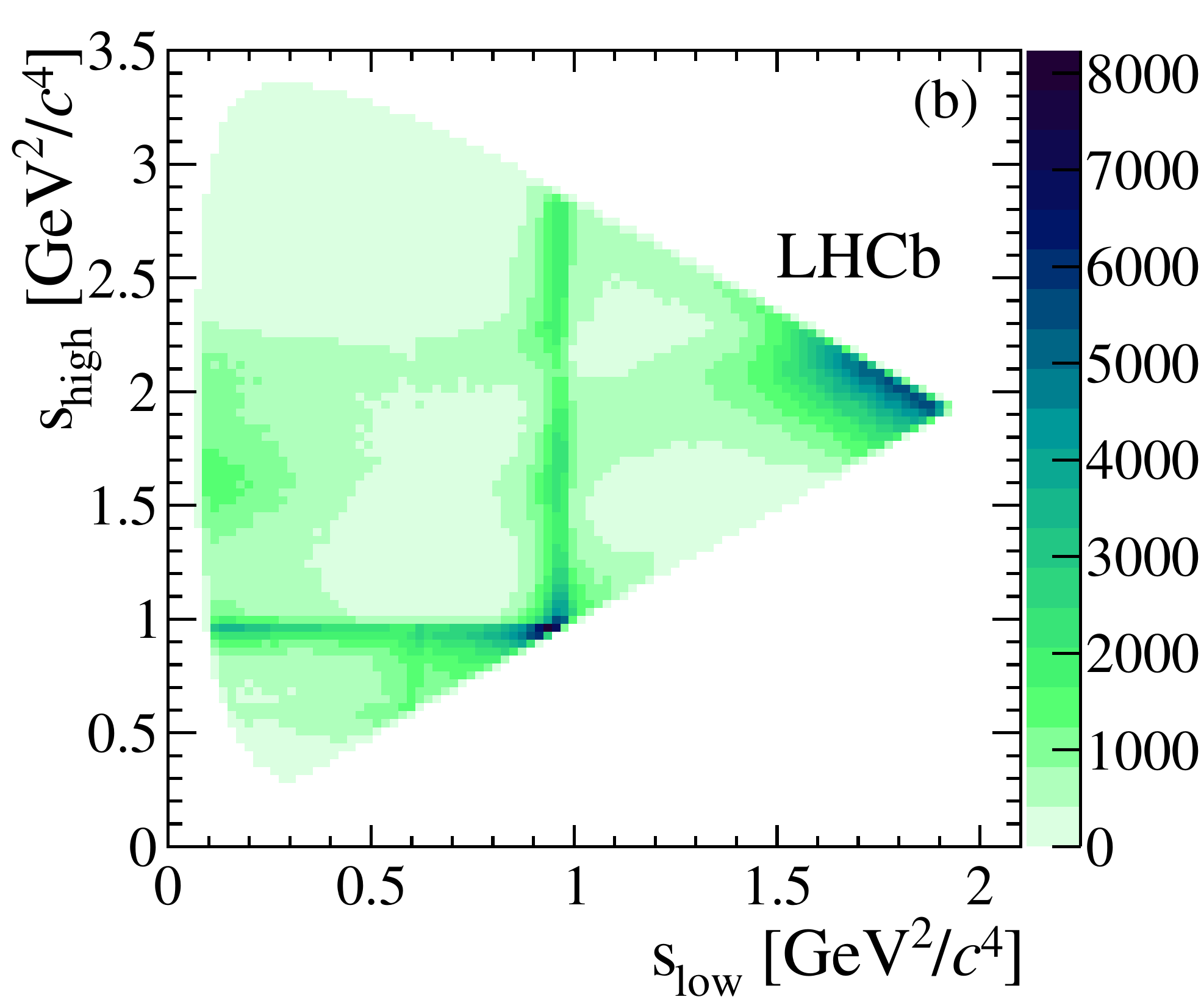} \\
\vspace*{-.3cm}
\caption{\small Dalitz plots for (a) \Dppp  and (b) \Dsppp  candidates selected within 
 $\pm2\tilde\sigma$ around the respective  $\tilde m$ weighted average mass.}
\end{center}
\label{fig:DpDp}
\end{figure}

The  \Dppp and \Dsppp Dalitz plots  are shown in Fig.~\ref{fig:DpDp},  with $s_{\rm low}$ and $s_{\rm high}$ being the lowest and highest  invariant mass squared combination, $M^2(\pi^-\pi^+)$,  respectively. Clear resonant structures are observed in both decay modes.

\section{Binned analysis}
\label{sec:binned}

\subsection{Method}

The binned method used to search for localised asymmetries in the \decay{\Dp}{\pim\pip\pip} decay phase space is based on a bin-by-bin comparison between the \Dp and \Dm 
 Dalitz plots~\cite{miranda, babarchi2}. For each bin of the Dalitz plot, the significance of the difference between the number of \Dp and \Dm candidates, \SCPi, is computed as

\begin{equation}
\SCPi \equiv  \frac{N_i^+-\alpha N_i^-}
{\sqrt{\alpha(N_i^++N_i^-)}} \ , \hskip .5cm \alpha \equiv\frac{ N^+}{N^-}, 
\label{eq:scp1}
\end{equation}
where
$N_i^+$ ($N_i^-$) is  the number of \Dp(\Dm) candidates in the $i\rm{th}$ bin and 
$N^+$ ($N^-$) is the sum of $N_i^+$ ($N_i^-$) over all  bins.
The parameter $\alpha$ removes the contribution of global asymmetries which may arise due to production~\cite{LHCb-PAPER-2012-026,LHCb-PAPER-2012-009} and detection asymmetries, as well as from \CPV. 
 Two binning schemes are used,  a uniform grid with bins of equal size and an adaptive binning
where the bins have the same population.

In the absence of localised asymmetries, the \SCPi values follow
 a standard normal Gaussian distribution. Therefore,  \CPV can be 
detected as a deviation from this behaviour. The numerical comparison between the \Dp and \Dm Dalitz plots is made by a $\chi^2$ test, with $\chi^2=\sum_{i} (\mathcal{S}_{\CP}^i)^2$. A p-value for the hypothesis of no \CPV is  obtained considering that the number of degrees of freedom (ndf) is equal to the total number of bins minus one, due to the constraint on the
overall \Dp/\Dm normalisation. 

A  \CPV signal is established  if a p-value lower than $3\!\times\! 10^{-7}$  is found, in which case it can be converted to a significance for the exclusion of
\CP symmetry in this channel.
 If no evidence of \CPV is found, this technique provides no model-independent way to set an upper limit. 

\subsection{Control mode and background}

\label{sec:background}

The search for local asymmetries across the \Dsppp Dalitz plot  is performed using both the uniform and the adaptive 
 (``\Ds adaptive'') binning schemes mentioned previously. A third scheme is also used: a ``scaled $D^+$" scheme, obtained  from the \Dp adaptive binning by scaling 
the bin edges by the ratios of the maximum values of $s_{\rm high}$(\Ds)/$s_{\rm high}$(\Dp) and $s_{\rm low}$(\Ds)/$s_{\rm low}$(\Dp). This scheme provides a one-to-one mapping of the corresponding Dalitz plots and allows  to probe regions in the signal and control channel phase spaces where  the momentum distributions of the three final state particles are similar.

\begin{figure}[htbp]
\begin{center}
\includegraphics*[width=0.46\textwidth]{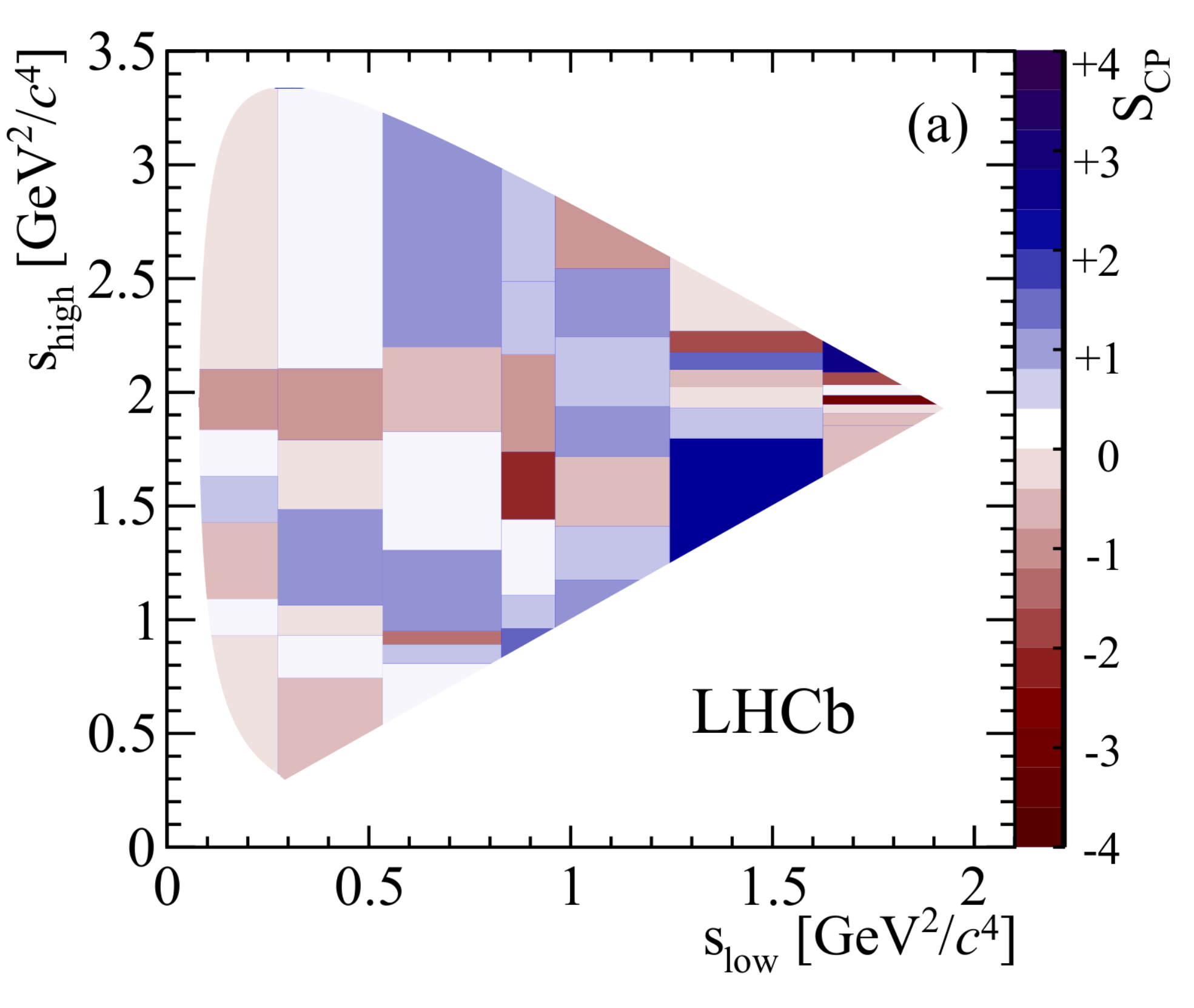}~~~
\includegraphics*[width=0.46\textwidth]{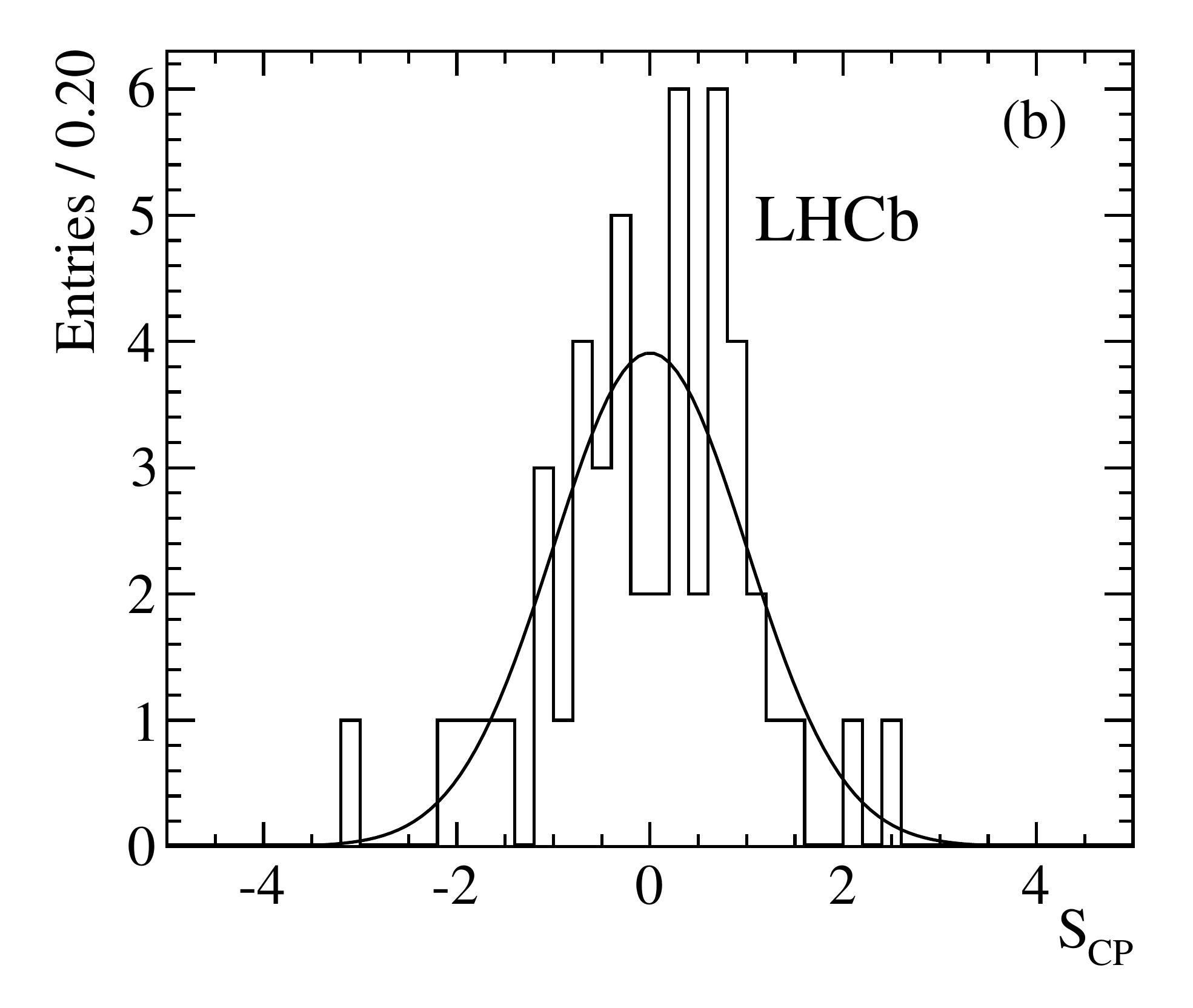}
\end{center}
\vspace*{-0.7cm}
\caption{\small (a) Distribution of $\mathcal{S}_{CP}^i$ with 49 \Ds adaptive bins of equal population in the \Dsppp Dalitz plot and (b) the corresponding one-dimensional distribution (histogram) with a standard normal Gaussian function superimposed  (solid line).}
\label{fig:MirandaDssyst1}
\end{figure}

The study is performed using $\alpha=0.992\pm 0.001$, as measured for the \Ds sample, and different granularities: 20, 30, 40, 49 and 100 adaptive bins for both the \Ds adaptive and scaled $D^+$ schemes, and 5$\times$5, 6$\times$7, 8$\times$9 and 12$\times$12 bins for the uniform grid scheme. Only bins with a minimum occupancy of 20 entries are considered. The p-values obtained are distributed in the range 4--87\%,  consistent with the hypothesis of absence of localised asymmetries. 
As an example, Fig.~\ref{fig:MirandaDssyst1} shows the distributions of \SCPi for the \Ds adaptive binning scheme with 49 bins. 

As a further cross-check, the \Ds sample is  divided according to magnet polarity and hardware trigger configurations. Typically, the p-values are  above 1\%, although one low value of 0.07\% is found for a particular trigger subset of magnet up data with 40 adaptive bins. When combined with magnet down data,  the p-value increases to 11\%.

The possibility of local asymmetries induced by the background under the \Dp signal peak is studied by considering the candidates with mass $M({\pi^-\pi^+\pi^+})$ in the ranges 1810--1835~\mevcc and 1905--1935~\mevcc, for which  $\alpha=1.000\pm 0.002$. 
Using an uniform grid with four different granularities, the p-values
are computed for each of the two sidebands.  The data are also divided according to the
magnet polarity.
The p-values are found to be within 0.4--95.5\%,  consistent with differences in the number of \Dp and \Dm candidates arising from statistical fluctuations. Since the selection criteria suppress  charm background decays to a negligible level, it is assumed that the background contribution to  the signal is similar to the sidebands. Therefore,  asymmetries eventually observed in the signal mode cannot be attributed to the background.

\subsection{Sensitivity studies}

\label{sec:binned.e791}

To study  the  \CPV sensitivity of the method for the current data set, a number of simulated pseudo-experiments are performed with sample size and purity similar to that observed in data. 
The  $D^+ \to \pi^-\pi^+\pi^+$ decays are generated according to an amplitude  model inspired by E791 results \cite{e791}, where the most important contributions originate from $\rho^0(770)\pi^+$, $\sigma(500)\pi^+$ and $f_2(1270)\pi^+$ resonant modes.  Background events are generated  evenly in the Dalitz plot. 
Since no theoretical predictions  on the presence or size of  \CPV are available for this channel, 
various scenarios are studied by introducing phase and magnitude differences between the main resonant modes for \Dp and \Dm. 
The sensitivity  for different binning strategies is also evaluated. 

Phase differences in the range $0.5$--$4.0^\circ$ and magnitude differences in the range $0.5$--$4.0\%$ are tested for  $\rho^0(770)\pi^+$, 
$\sigma(500)\pi^+$  and $f_2(1270)\pi^+$ modes. The study  shows a sensitivity (p-values below $10^{-7}$) around $1^\circ$ to $2^\circ$ in phase differences and $2\%$ in amplitude in these channels. The sensitivity decreases when the number of bins is larger than 100, so a few tens of bins approaches the optimal choice. A slightly better sensitivity for the adaptive binning strategy  is found in most of the studies.

Since the presence of background tends to dilute a potential sign of \CPV, additional pseudo-experiment studies are made for different 
scenarios  based on signal yields and purities attainable on data. Results show that better sensitivities are found for higher yields, despite the lower purity.

\section{Unbinned analysis}
\label{sec:unbinned}

\subsection{k-nearest neighbour analysis technique}
\label{sec:knntechnique}

The unbinned model-independent method 
of searching for \CPV in many-body decays uses 
the concept of nearest neighbour events 
in a combined $D^+$ and $D^-$ samples to test whether they share 
the same parent distribution function~\cite{Williams:2010vh,henze,schilling}.
To find the $n_k$ nearest neighbour events of each $D^+$ and $D^-$ event, 
the Euclidean distance between points in the Dalitz plot 
of three-body $D^+$ and $D^-$ decays is used.  
For the whole event sample
a test statistic $T$ for the null hypothesis is calculated,
\begin{equation}
T = \frac{1}{n_k(N_++N_-)}\sum\limits_{i=1}^{N_++N_-}\sum\limits_{k=1}^{n_k}I(i,k),
\end{equation}
where  $I(i,k)=1$ if the $i$th event and its $k$th nearest neighbour have
the same charge
and $I(i,k)=0$ otherwise and $N_+$ ($N_-$) is the number of events
in the $D^+$ ($D^-$) sample.

The test statistic $T$ is the mean fraction of like-charged neighbour pairs
in the combined $D^+$ and $D^-$ decays sample. The advantage of 
the k-nearest neighbour method  (kNN), in comparison with other proposed 
methods for unbinned analyses~\cite{Williams:2010vh}, is that 
the calculation of $T$ is simple and fast 
and the expected distribution of $T$ is well known: for the null hypothesis 
it follows a Gaussian distribution with mean $\mu_T$ and 
variance $\sigma^2_T$  calculated from known parameters of the distributions,
\begin{equation}
\mu_T = \frac{N_+(N_+-1)+N_-(N_--1)}{N(N-1)},
\label{eq:mut}
\end{equation}
\begin{equation}
\lim_{N,n_k,D \to \infty} \sigma_T^2 = \frac{1}{Nn_k}\left(\frac{N_+N_-}{N^2}+4\frac{N_+^2N_-^2}{N^4}\right),
\label{eq:sigma}
\end{equation}
where $N=N_++N_-$ and $D$ is a space dimension.  
For $N_+ = N_-$ a reference value
\begin{equation}
\mu_{\it TR} = \frac{1}{2} \left(\frac{N-2}{N-1}\right)
\label{eq:muthalf}
\end{equation}
is obtained and for a very large number of events $N$,
$\mu_T$ approaches $0.5$. However, since the observed deviations of $\mu_T$
from $\mu_{\it TR}$ are sometimes tiny, it is necessary to calculate
$\mu_T-\mu_{\it TR}$.
The convergence in Eq.~\ref{eq:sigma} is fast and $\sigma_T$ can be obtained with 
a good approximation even for space dimension $D=2$ for the current values 
of $N_+$, $N_-$ and $n_k$~\cite{Williams:2010vh,schilling}.

The kNN method is applied to search for \CPV
in a given region of the Dalitz plot in two ways:
by looking at a ``normalisation'' asymmetry ($N_+ \neq N_-$ in a given region)
using a pull $(\mu_T-\mu_{\it TR})/\Delta(\mu_T-\mu_{\it TR})$ variable, where
the uncertainty on $\mu_T$ is $\Delta\mu_T$
and the uncertainty on $\mu_{\it TR}$ is $\Delta\mu_{\it TR}$, and looking
for a ``shape'' or particle density function (pdf) asymmetry using another pull $(T-\mu_T)/\sigma_T$ variable.

As in the binned method, this technique provides no model-independent way to set an upper limit if no \CPV is found.


\subsection{Control mode and background}

The Cabibbo-favoured $D^+_s$ decays serve as a control sample to estimate
the size of production and detection asymmetries and systematic effects.
The sensitivity for local \CPV in the Dalitz plot of the kNN method 
can be increased by taking into account 
only events from the region where \CPV is expected to be enhanced 
by the known intermediate resonances in the decays.
Since these regions are  characterised by 
enhanced variations of strong phases, the conditions for observation
of \CPV are more favourable. 
Events from  other regions are expected to only dilute the signal of \CPV.

\begin{figure}[htbp]
\begin{center}
\includegraphics*[width=0.45\textwidth]{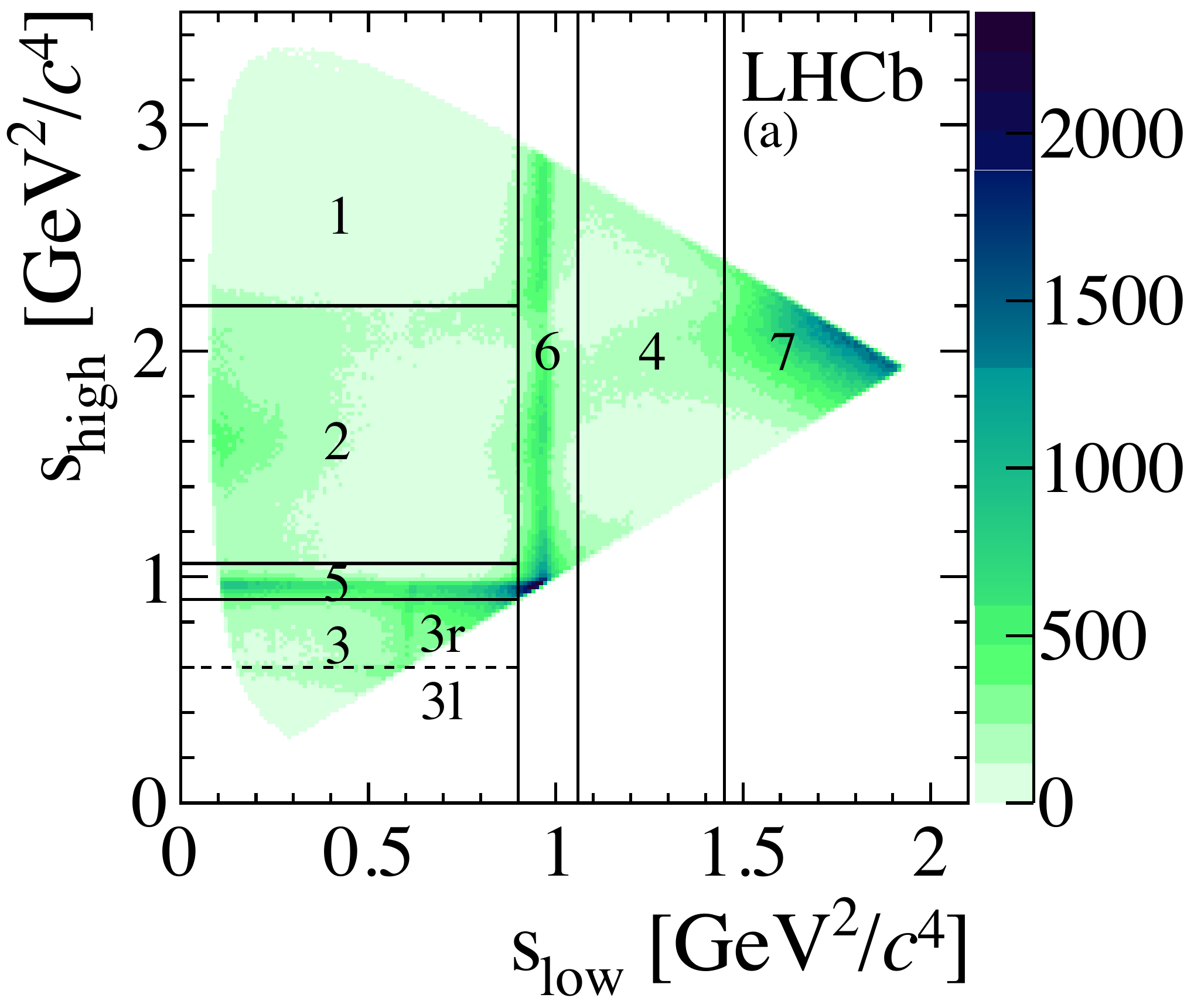}~~~
\includegraphics*[width=0.45\textwidth]{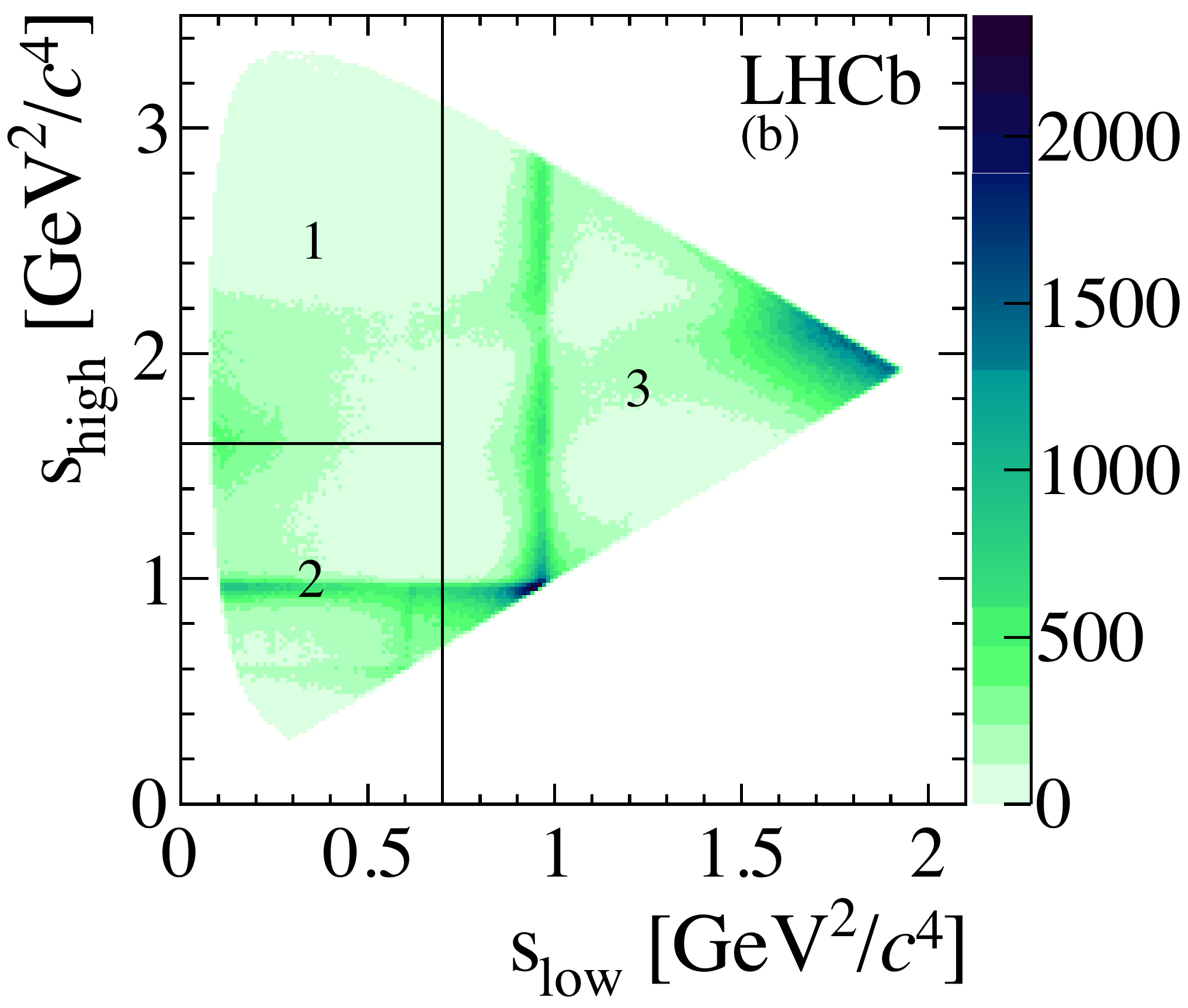} \\
\end{center}
\vspace*{-0.7cm}
\caption{\small Dalitz plot for \Dsppp control sample decays divided into
              (a) seven regions R1-R7 and 
              (b) three regions P1-P3. 
              Region R3 is further divided into two regions
              of $s_{\rm high}$ at masses smaller  (R3l) 
              and larger (R3r) than the $\rho^{0}(770)$ resonance.}
  \label{fig:Dsregions}
\end{figure}

The Dalitz plot for the control channel \Dsppp  is partitioned into three (P1-P3)
or seven (R1-R7) regions shown in Fig.~\ref{fig:Dsregions}. 
The division R1-R7 is such that regions enriched in resonances
are separated from regions dominated by smoother distributions of events. 
Region R3 is further divided into two regions  of $s_{\rm high}$
at masses smaller (R3l) 
and larger (R3r) than the $\rho^{0}(770)$ resonance, in order  to study 
possible asymmetries due to a sign change of the strong phase 
when crossing the resonance pole. 
The three regions P1-P3 correspond to 
a more complicated structure of resonances in  the signal decay \Dppp
(see Fig.~\ref{fig:Dregions}).

\begin{figure}[htbp]
\begin{center}
\includegraphics*[width=0.45\textwidth]{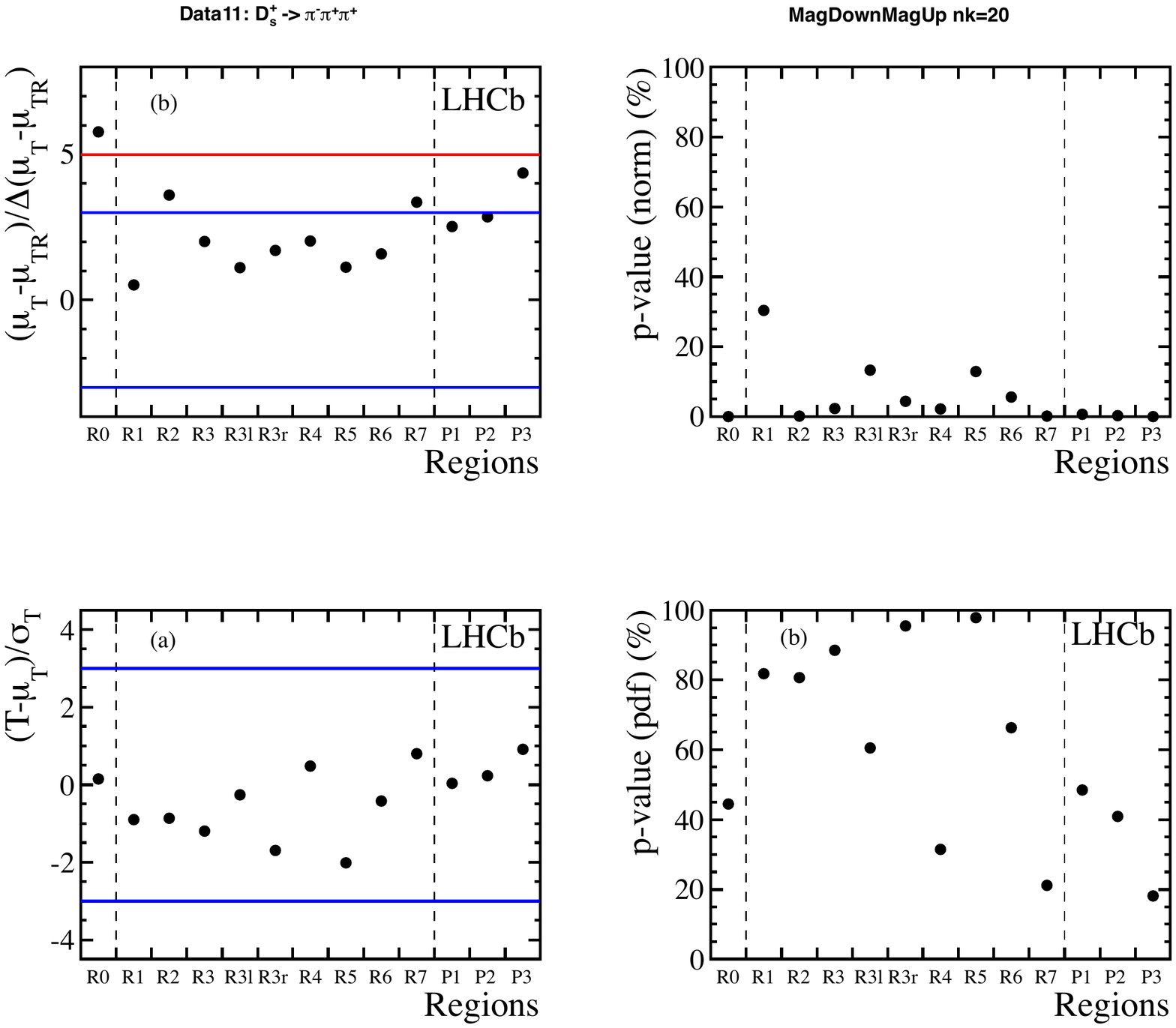}~~~
\includegraphics*[width=0.45\textwidth]{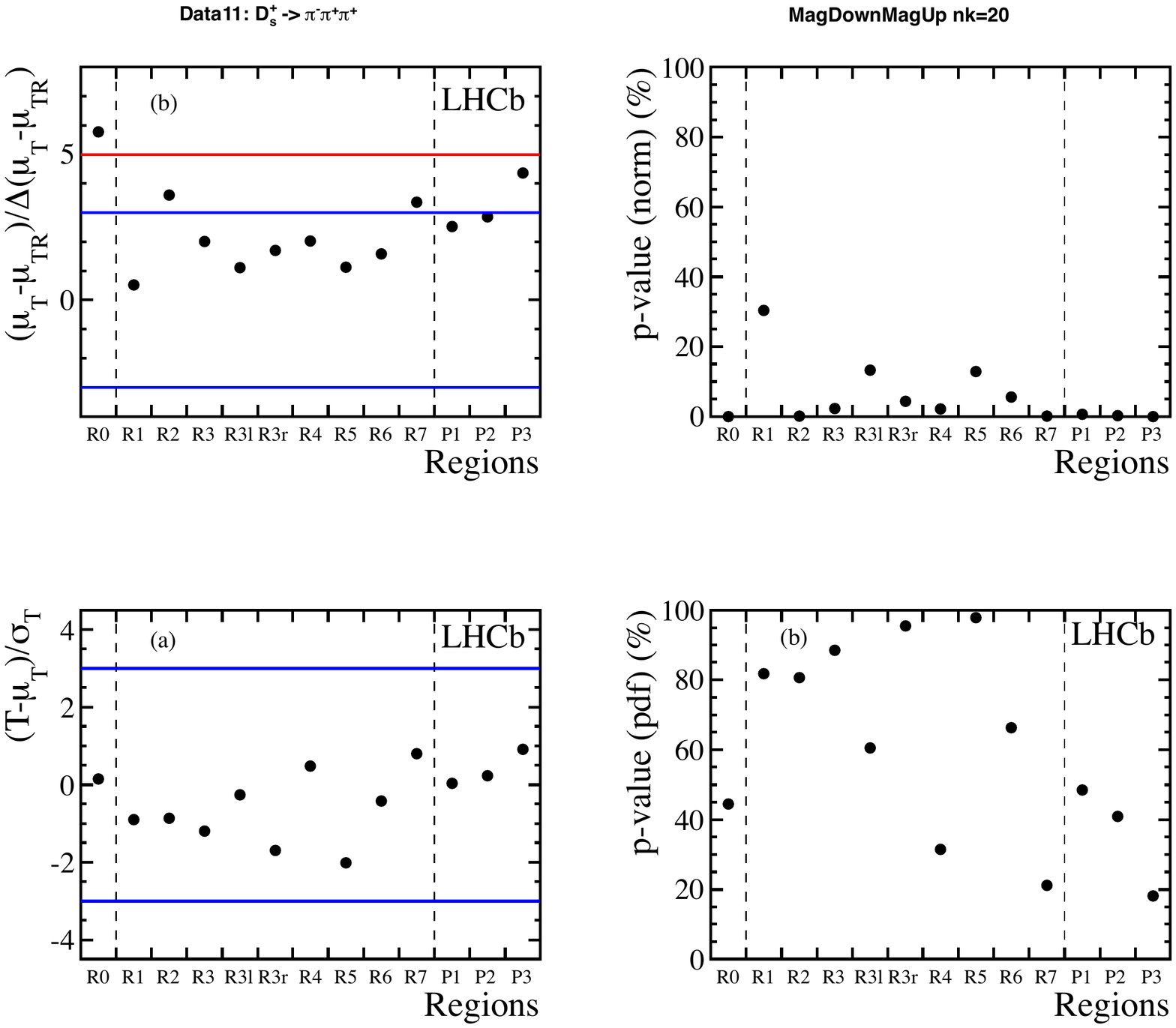}\\
\end{center}
\vspace*{-0.7cm}
\caption{\small (a) Pull values of $T$ and (b)  the corresponding p-values
              for \Dsppp control sample candidates restricted to each region,
              obtained using the kNN method with $n_k=20$. 
              The horizontal blue lines in (a) represent $-3$ and $+3$ pull values. 
              The region R0 corresponds to the full Dalitz plot.
              Note that the points for the overlapping regions are correlated.}
  \label{fig:DsMagDownMagUp_nk20}
\end{figure}

\begin{figure}[htbp]
\begin{center}
\includegraphics*[width=0.45\textwidth]{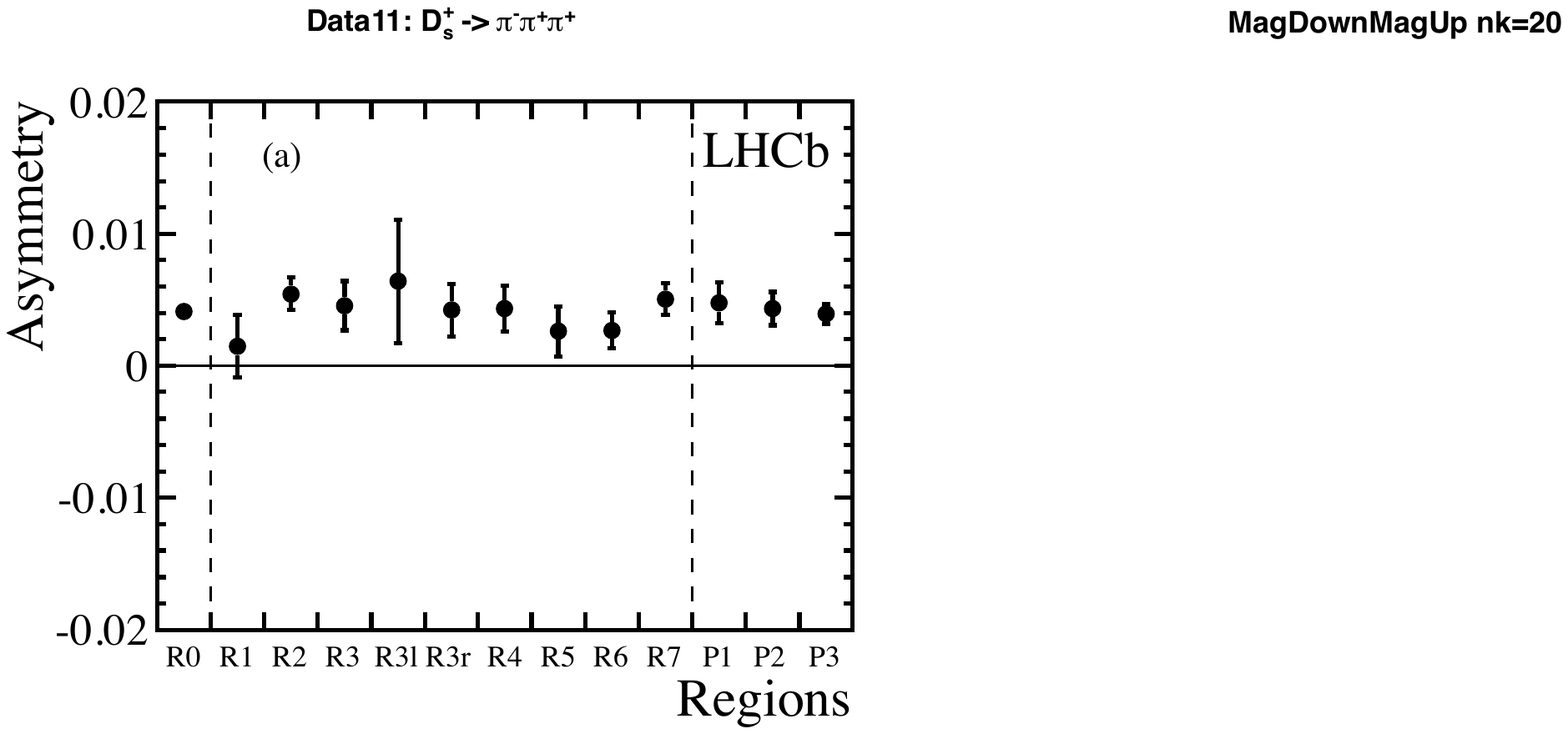}~~~
\includegraphics*[width=0.44\textwidth]{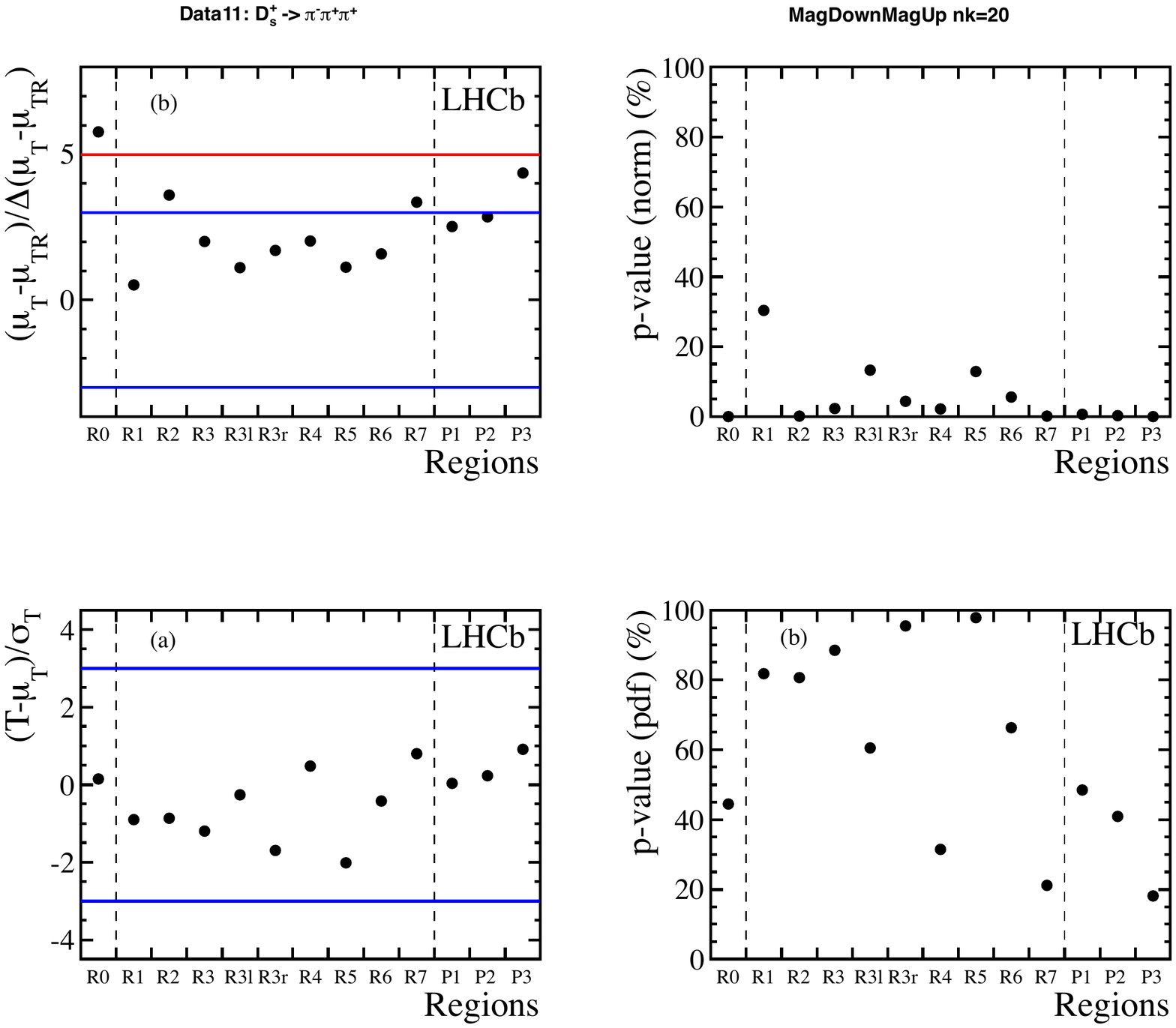}\\
\end{center}
\vspace*{-0.7cm}
\caption{\small (a) Raw asymmetry $A=(N_--N_+)/(N_-+N_+)$ and (b) the pull values of 
              $\mu_T$ for \Dsppp control sample candidates restricted to each region.
              The horizontal lines in (b) represent $+3$
              and $+5$ pull values. 
              The region R0 corresponds to the full Dalitz plot.
              Note that the points for the overlapping regions are correlated.}
  \label{fig:DsMagDownMagUp_asym}
\end{figure}

The value of the test statistic $T$ measured using the kNN method with $n_k=20$
for the full Dalitz plot (called R0) of
\Dsppp candidates 
is compared to the expected Gaussian $T$ distribution
with $\mu_T$ and $\sigma_T$ calculated from data. 
The calculated p-value is 44\% for the hypothesis 
of no \CP asymmetry.  The p-values are obtained by integrating the
Gaussian $T$ distribution from a given value up to its maximum value of 1.
The results are shown 
in Fig.~\ref{fig:DsMagDownMagUp_nk20} separately for each region. 
They do not show any asymmetry between $D_s^+$ and $D_s^-$ samples. 
Since no \CPV is expected in the control channel,
the local detection asymmetries are smaller than the present sensitivity of the kNN method.
The production asymmetry is accounted for
in the kNN method as a deviation of the measured value of $\mu_T$
from the reference value $\mu_{\it TR}$. In the present sample, the obtained 
value $\mu_T-0.5=(84\pm 15)\times 10^{-7}$,
with $(\mu_T-\mu_{\it TR})/\Delta(\mu_T-\mu_{\it TR})=5.8\sigma$,
in the full Dalitz plot is a consequence of the observed global asymmetry
of about 0.4\%.  This value is consistent with the previous measurement 
from LHCb~\cite{LHCb-PAPER-2012-009}.
The comparison of the raw asymmetry $A=(N_--N_+)/(N_-+N_+)$
and the pull values of $\mu_T$ in all regions are presented
in Fig.~\ref{fig:DsMagDownMagUp_asym}. The measured raw asymmetry
is similar in all regions as expected for an effect due to 
the production asymmetry. It is  interesting to note
the relation $\mu_T-\mu_{\it TR}\approx A^2/2$ at order $1/N$
between the raw asymmetry and the parameters of the kNN method.

A region-by-region comparison of $D_s^+$ candidates for magnet down and magnet up data
gives insight into left-right detection asymmetries.
No further asymmetries,
except for the global production asymmetry discussed above, are found. 

The number of nearest neighbour events $n_k$  is the only 
parameter of the kNN method. 
The results for the control channel show no significant dependence 
of p-values on $n_k$. 
Higher values of $n_k$ reduce statistical fluctuations due to the local population 
density and should be preferred. On the other hand, increasing the number 
of nearest neighbours with limited number of events in the sample can quickly increase 
the radius of the local region under investigation.

\begin{figure}[tbp]
\begin{center}
\includegraphics*[width=0.45\textwidth]{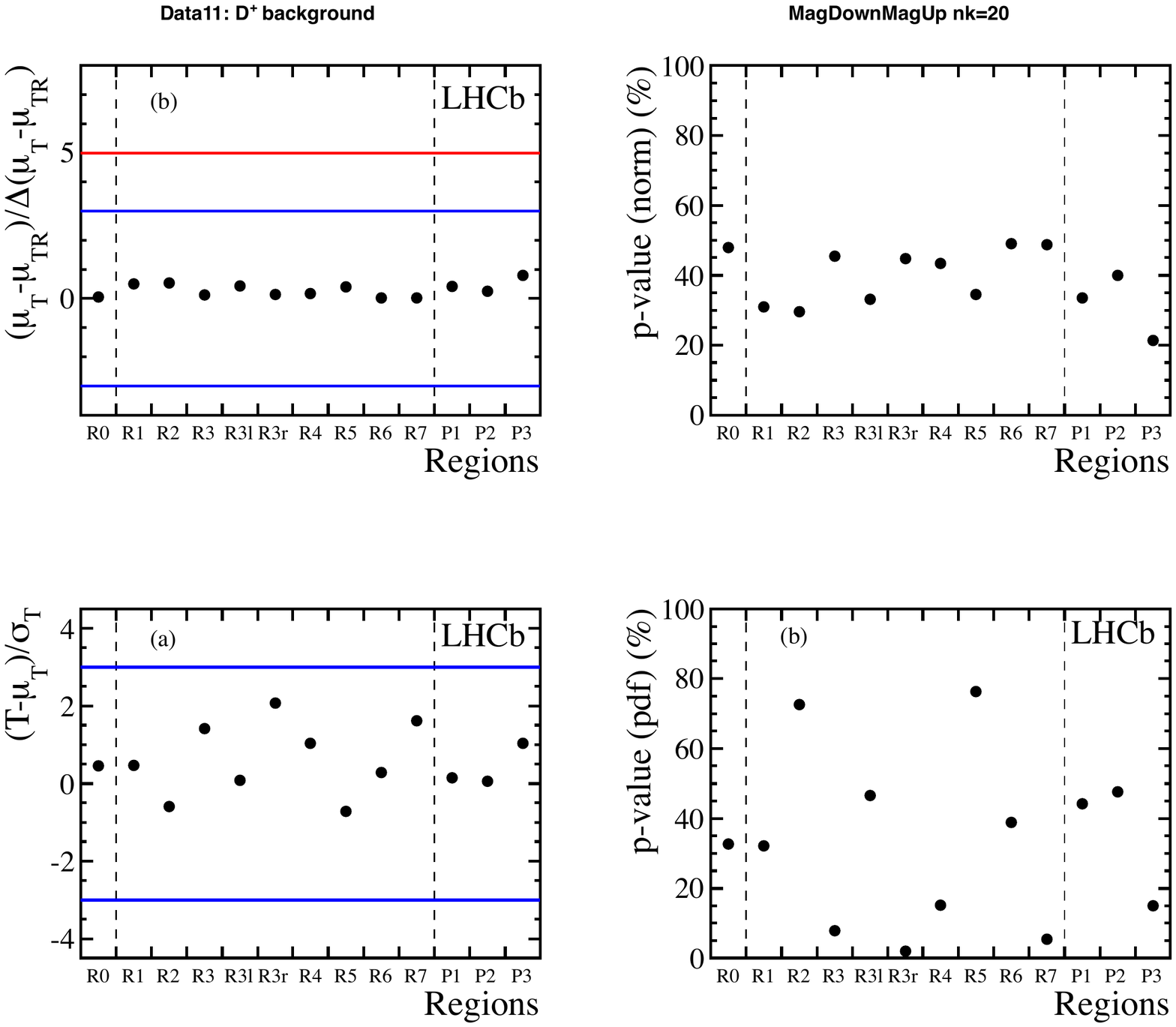}~~~
\includegraphics*[width=0.46\textwidth]{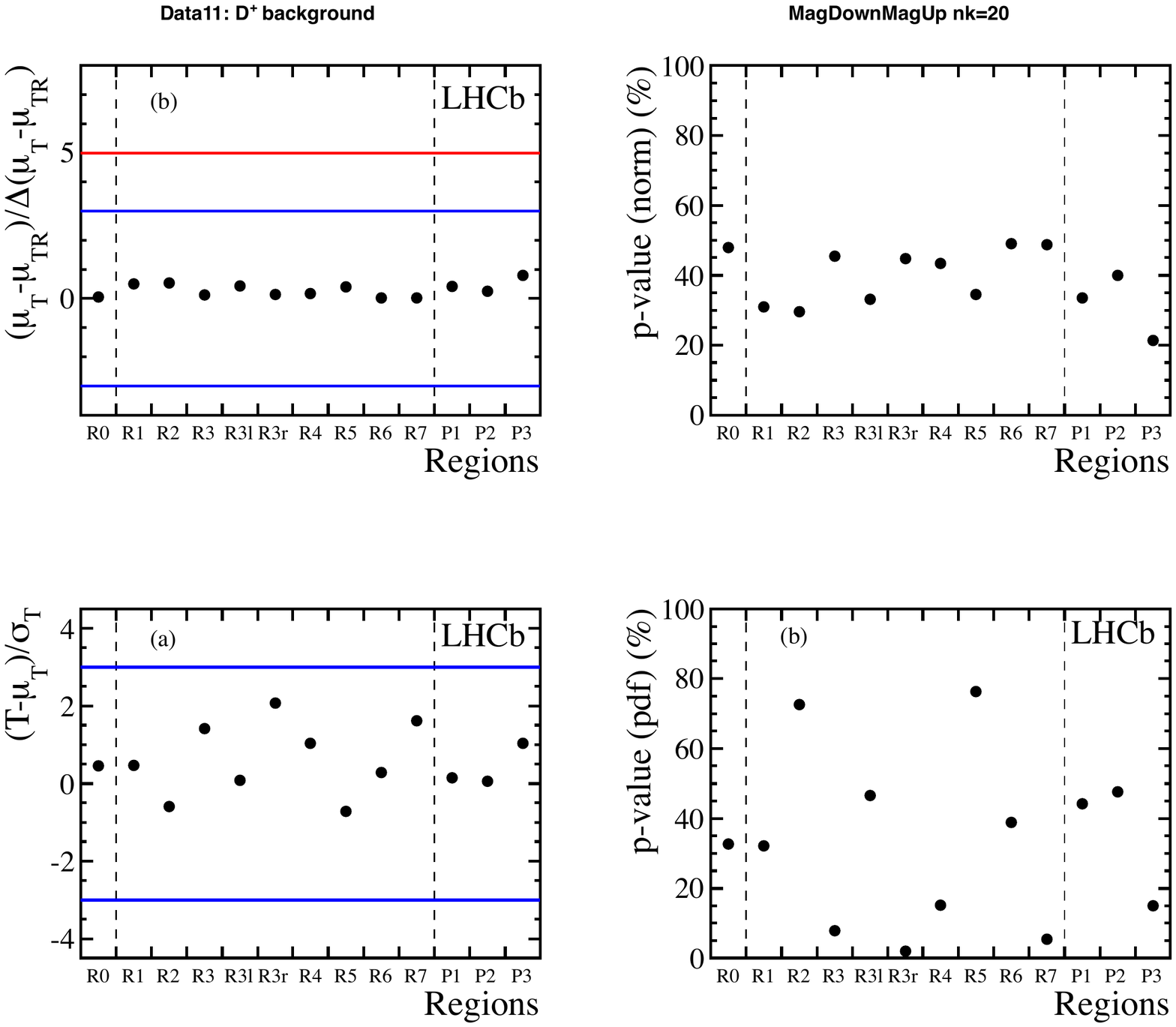} \\
\end{center}
\vspace*{-0.7cm}
\caption{\small (a) Pull values of $T$ and (b) the corresponding p-values
              for the background candidates restricted to each region 
              obtained using the kNN method with $n_k=20$. 
              The horizontal blue lines in (a) represent $-3$ and $+3$ 
              pull values. The region R0 corresponds to the full Dalitz plot.
              Note that the points for the overlapping regions are correlated.}
  \label{fig:DbgdMagDownMagUp_nk20}
\end{figure}

\begin{figure}[tbp]
\begin{center}
\includegraphics*[width=0.46\textwidth]{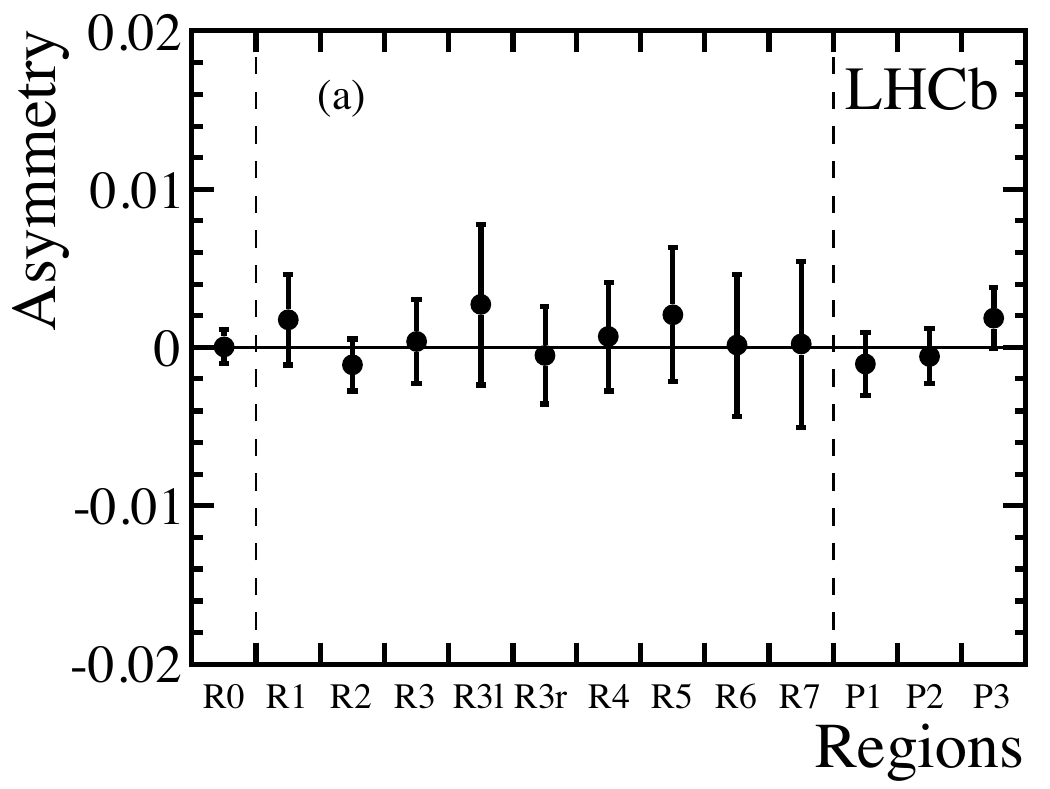}~~~
\includegraphics*[width=0.445\textwidth]{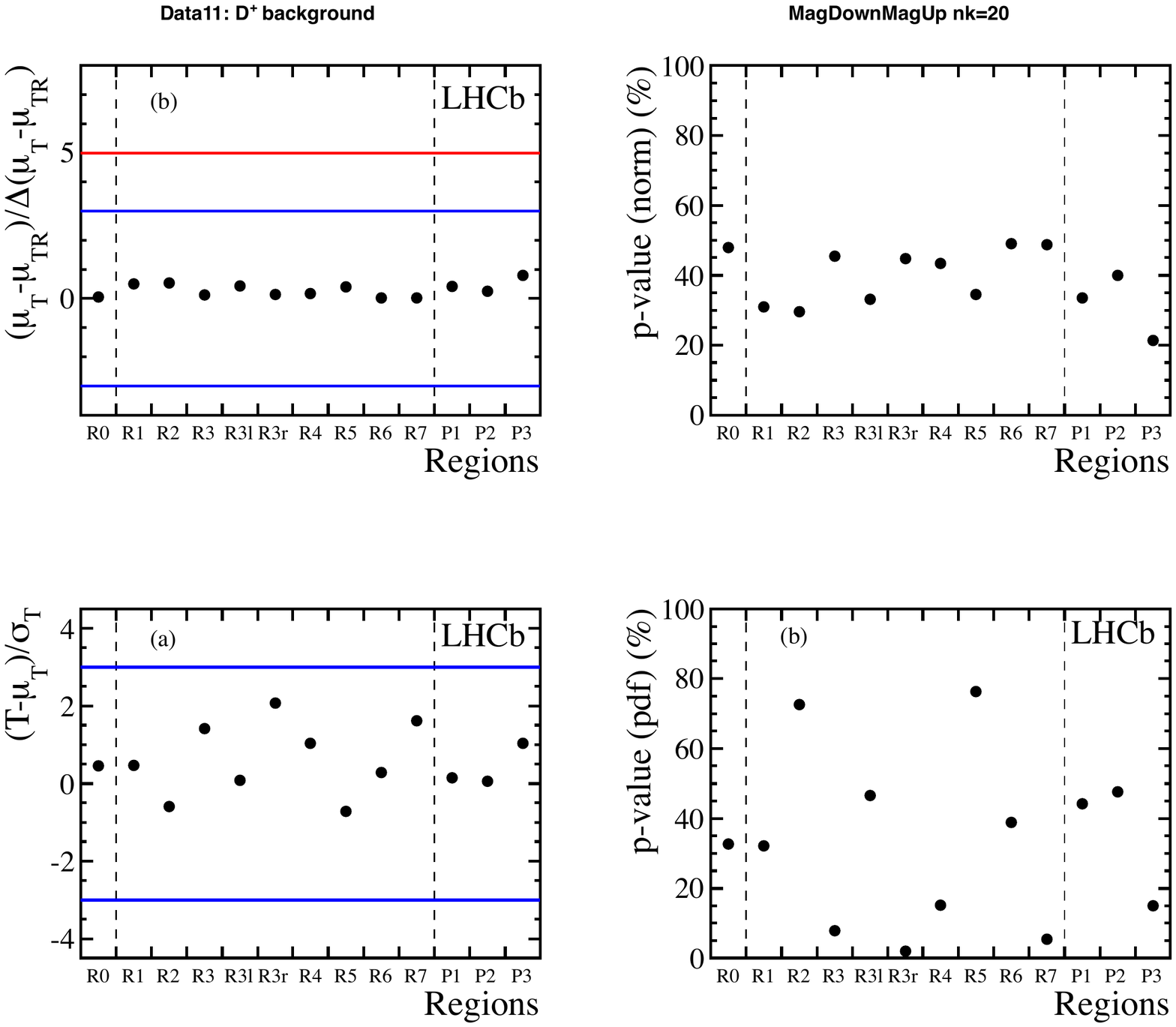} \\
\end{center}
\vspace*{-0.7cm}
\caption{\small (a) Raw asymmetry and (b) pull value of 
              $\mu_T$ as a function of a region
              for the background candidates. The horizontal lines in (b) represent $+3$
              and $+5$ pull values. 
              The region R0 corresponds to the full Dalitz plot.
              Note that the points for the overlapping regions are correlated.}
  \label{fig:DbgdMagDownMagUp_asym}
\end{figure}

The kNN method also is applied to the background events,
defined in Sec.~\ref{sec:background}.
Contrary to the measurements for the \Dsppp candidates,
for background no production asymmetry is observed.
The calculated $\mu_T-0.5=(-5.80\pm 0.46)\times 10^{-7}$ 
for the full Dalitz plot is very close 
to the value $\mu_{\it TR}-0.5=(-5.8239\pm 0.0063)\times 10^{-7}$ expected 
for an equal number of events in $D^+$ and $D^-$ samples (Eq.~\ref{eq:muthalf}).
The measured pull values of $T$ and the corresponding p-values obtained using
the kNN method with $n_k=20$ are
presented for the background in Fig.~\ref{fig:DbgdMagDownMagUp_nk20}, 
separately for each region. 
The comparison of normalisation asymmetries and pull values of $\mu_T$ in all regions are
presented in Fig.~\ref{fig:DbgdMagDownMagUp_asym}. All the kNN method results 
are consistent with no significant asymmetry.

\subsection{Sensitivity studies}

The sensitivity of the kNN method  is tested with the same pseudo-experiment model
described in  Sec.~\ref{sec:binned.e791}.
If the simulated asymmetries are spread out in the Dalitz plot the events may be
moved from one region to another.
For these asymmetries it is observed that the difference 
in shape of the probability density functions
is in large part absorbed in the difference in the normalisation. 
This indicates that the choice of the regions is important for increasing 
the sensitivity of the kNN method. 
In general the method applied in a given 
region is sensitive to weak phase 
differences greater than $(1-2)^{\circ}$ and  magnitude
differences of $(2-4)$\%.


\section{Results}
\label{sec:results}

\subsection{Binned method}

The search for \CPV in the Cabibbo-suppressed mode  \Dppp is pursued following the strategy described in Section~\ref{sec:binned}.
For the total sample size of about 3.1~million \Dp and \Dm candidates, the normalisation factor $\alpha$, defined in Eq.~\ref{eq:scp1},  is $0.990 \pm 0.001$.
Both adaptive and uniform binning schemes in the Dalitz plot are used for different binning sizes. 

\begin{figure}[htbp]
\begin{center}
\includegraphics*[width=0.46\textwidth]{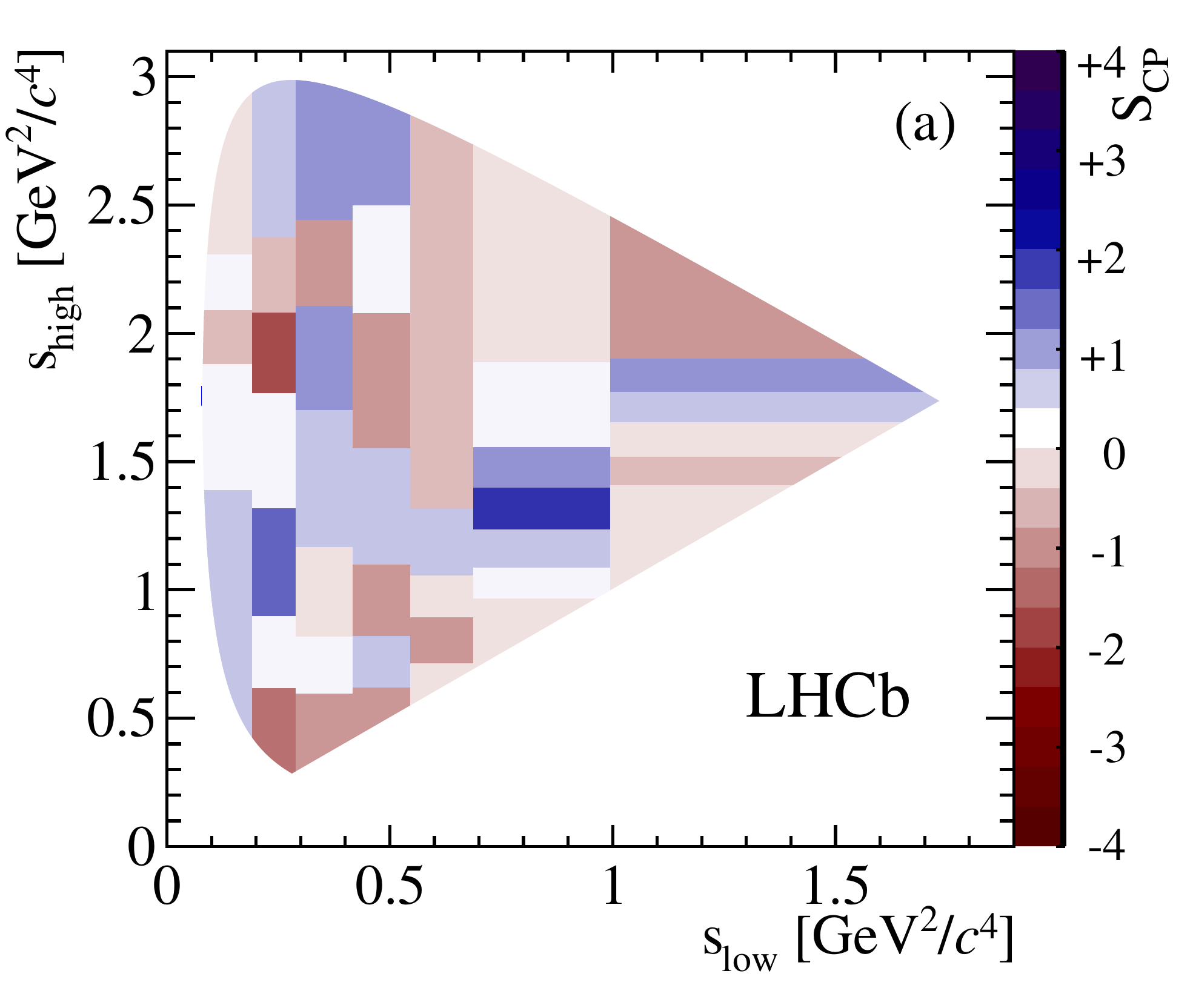}
\includegraphics*[width=0.46\textwidth]{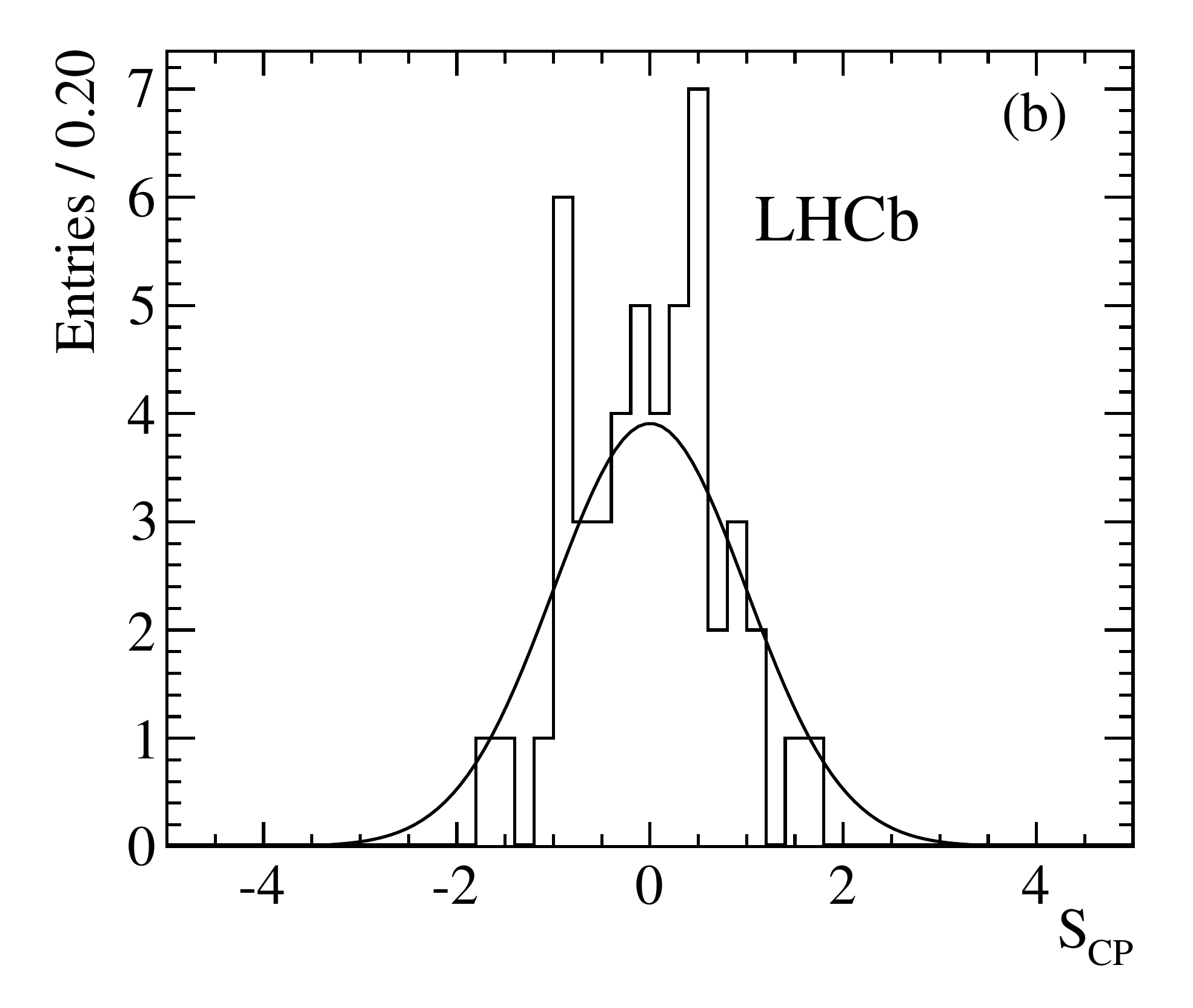}
\includegraphics*[width=0.46\textwidth]{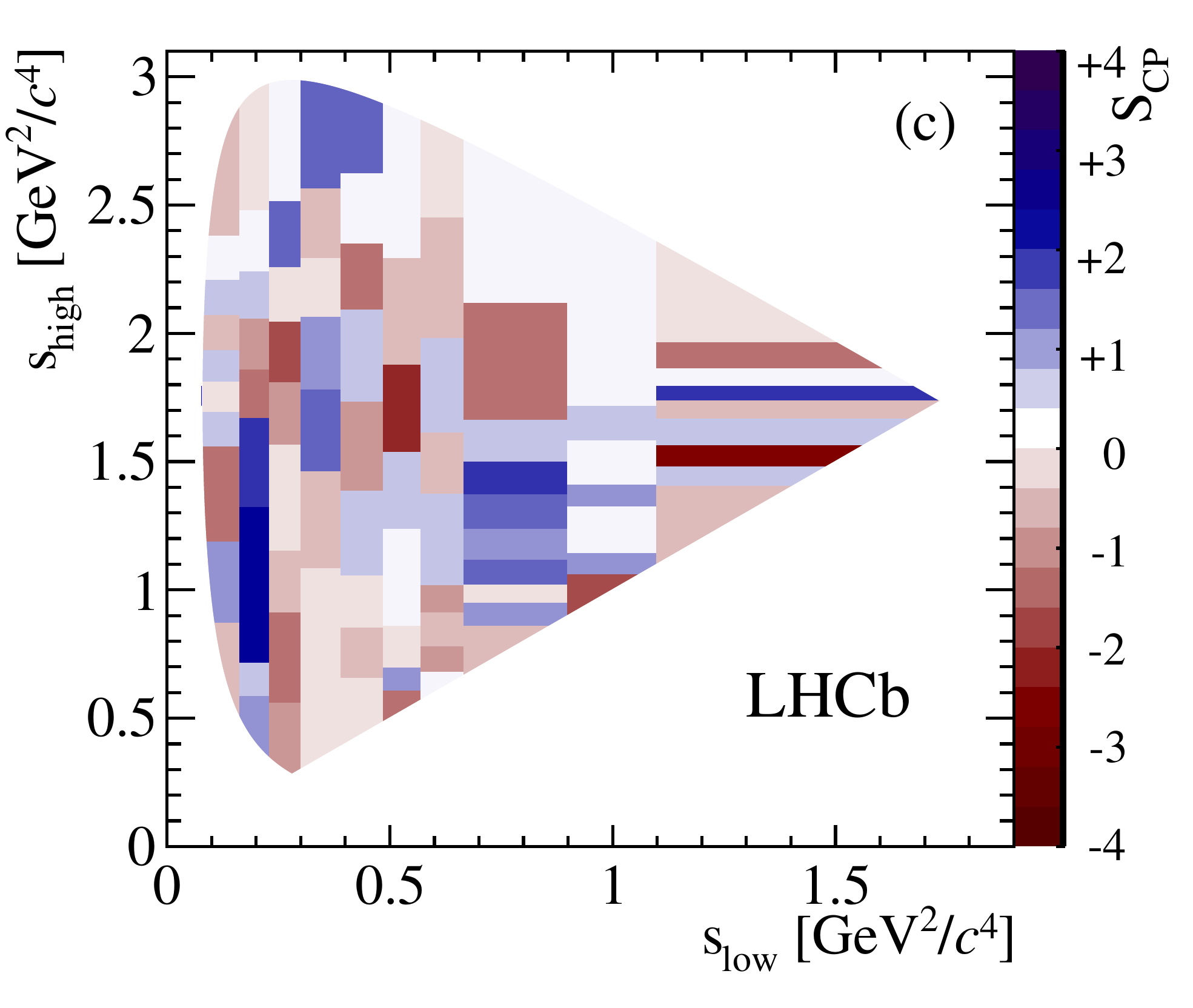}
\includegraphics*[width=0.46\textwidth]{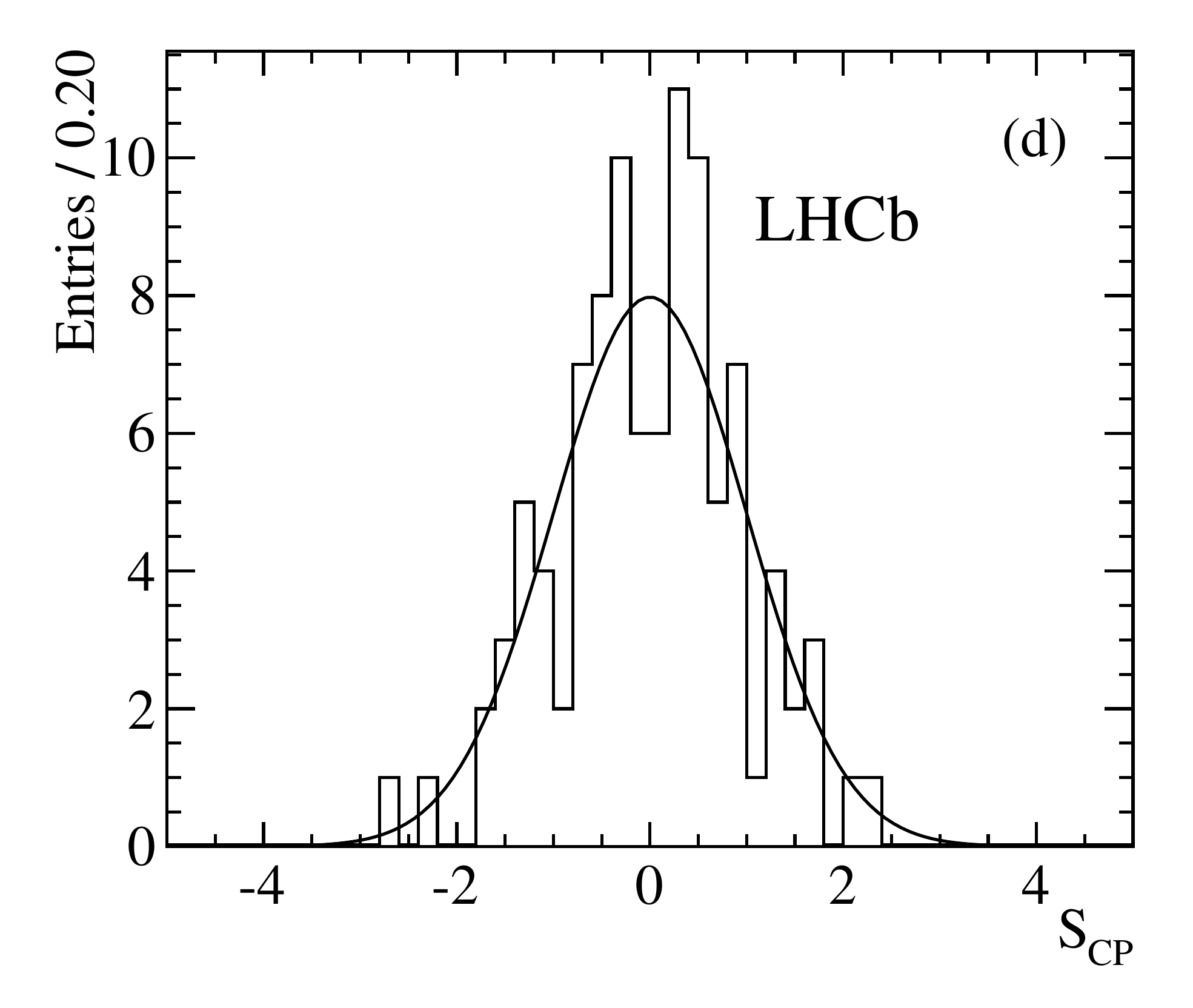}

\end{center}
\vspace*{-.7cm}

\caption{\small Distributions of $\mathcal{S}_{CP}^i$ across the \Dp Dalitz plane, with the adaptive binning scheme of uniform population for the total \Dppp data sample with (a) 49 and (c) 100 bins. The corresponding one-dimensional $\mathcal{S}_{CP}^i$ distributions (b) and (d) are shown with a standard normal Gaussian function superimposed  (solid line).}
\label{fig:Dmiranda2D}
\end{figure}

The  $\mathcal{S}_{CP}^i$ values across  the Dalitz plot and the corresponding  histogram for the adaptive binning scheme with  49 and 100 bins are illustrated in Fig.~\ref{fig:Dmiranda2D}. The p-values for these and other binning choices are shown in Table~\ref{tab:resMirandaAdaptativeD}. All p-values show statistical agreement between the \Dp and \Dm samples. 

\begin{figure}[htpb]
\begin{center}
\includegraphics*[width=0.46\textwidth]{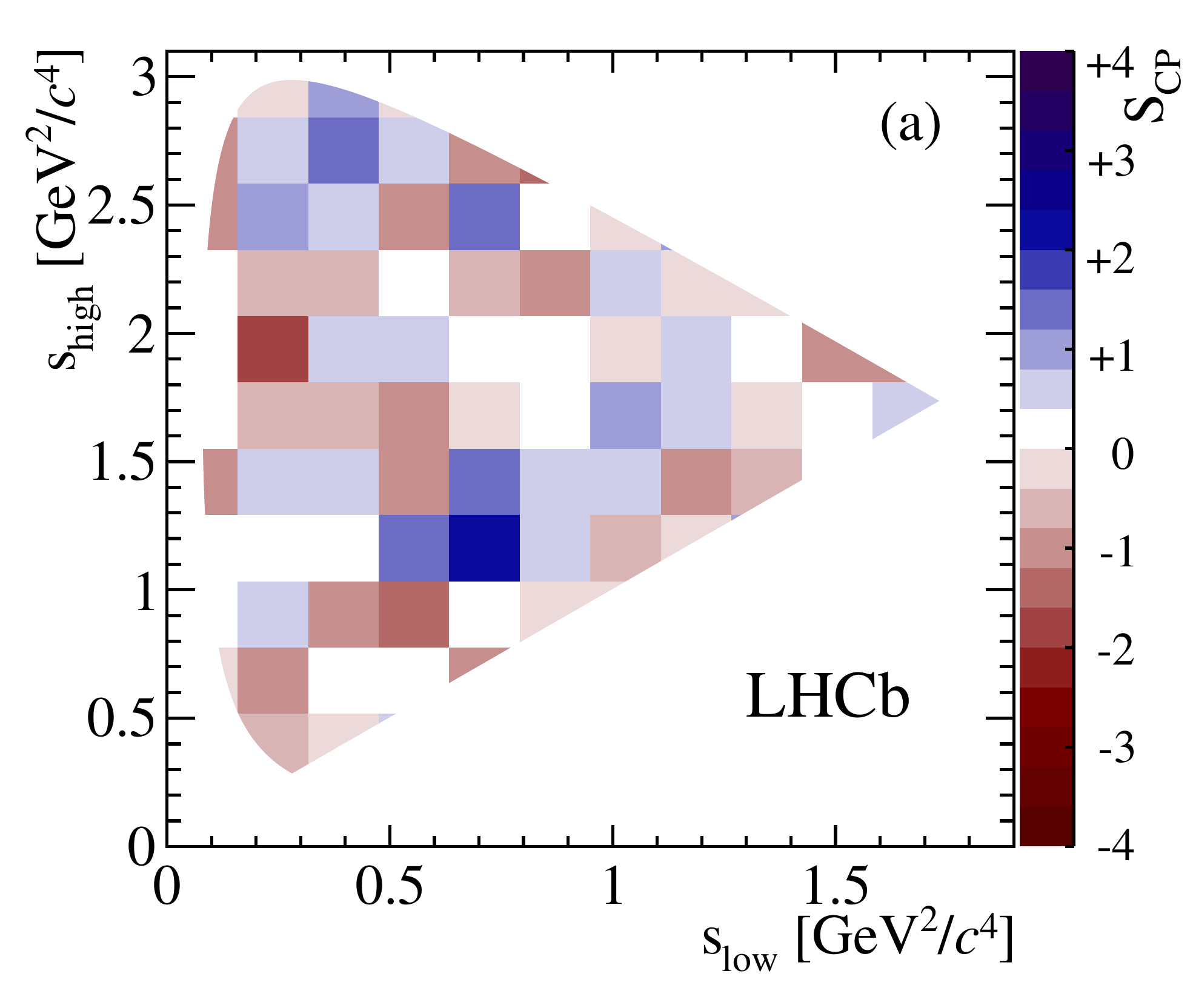}
\includegraphics*[width=0.46\textwidth]{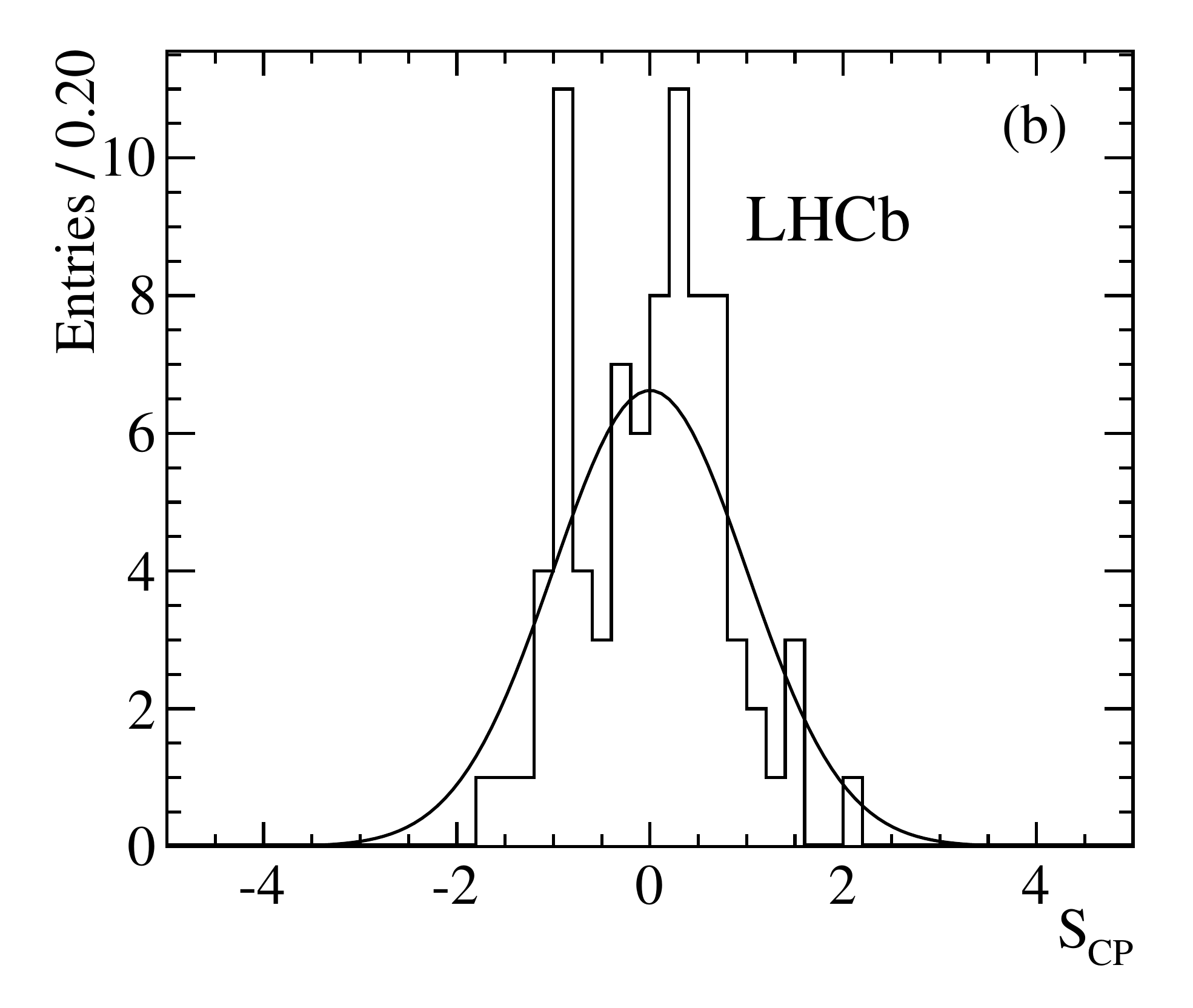}
\end{center}
\vspace*{-.7cm}

\caption{\small  (a) Distribution of $\mathcal{S}_{CP}^i$ with 98 bins in the uniform binning scheme for the total \Dppp data sample and (b) the corresponding one-dimensional $\mathcal{S}_{CP}^i$ distribution (b)  with a standard normal  Gaussian function superimposed (solid line). }
\label{fig:MirandaDsyst1}
\end{figure}

The same  $\chi^2$ test is performed for the uniform binning scheme, using 20, 32, 52 and 98 bins also resulting in p-values consistent with the null hypothesis, all above 90\%. The $\mathcal{S}_{CP}^i$ distribution in the Dalitz plot for 98 bins  and the corresponding histogram is shown in Fig.~\ref{fig:MirandaDsyst1}.

As consistency checks, the analysis is repeated with independent subsamples obtained by  separating the total sample according to magnet polarity, hardware trigger configurations, and data-taking periods. The resulting p-values  range from 0.3\% to 98.3\%. 

All the results above indicate the absence of \CPV in the  \Dppp channel at the current analysis sensitivity. 

\begin{table}[htbp]
\caption{\small Results for the $D^+ \to \pi^-\pi^+\pi^+$ decay sample using the adaptive binning scheme with different numbers of bins. 
The number of degrees of freedom is the number of bins minus 1.}
\begin{center}
 \begin{tabular}{ccc}\hline

Number of bins  &        $\chi^2$  & p-value (\%)         \\\hline
 20   & 14.0 & 78.1  \\
 30   & 28.2 & 50.6 \\
 40   & 28.5 & 89.2 \\
 49   & 26.7 & 99.5 \\
 100  & 89.1 & 75.1   \\\hline
\end{tabular}
\label{tab:resMirandaAdaptativeD}
\end{center}
\end{table}
 
\subsection{Unbinned method}

\begin{figure}[tbp]
\begin{center}
\includegraphics*[width=0.45\textwidth]{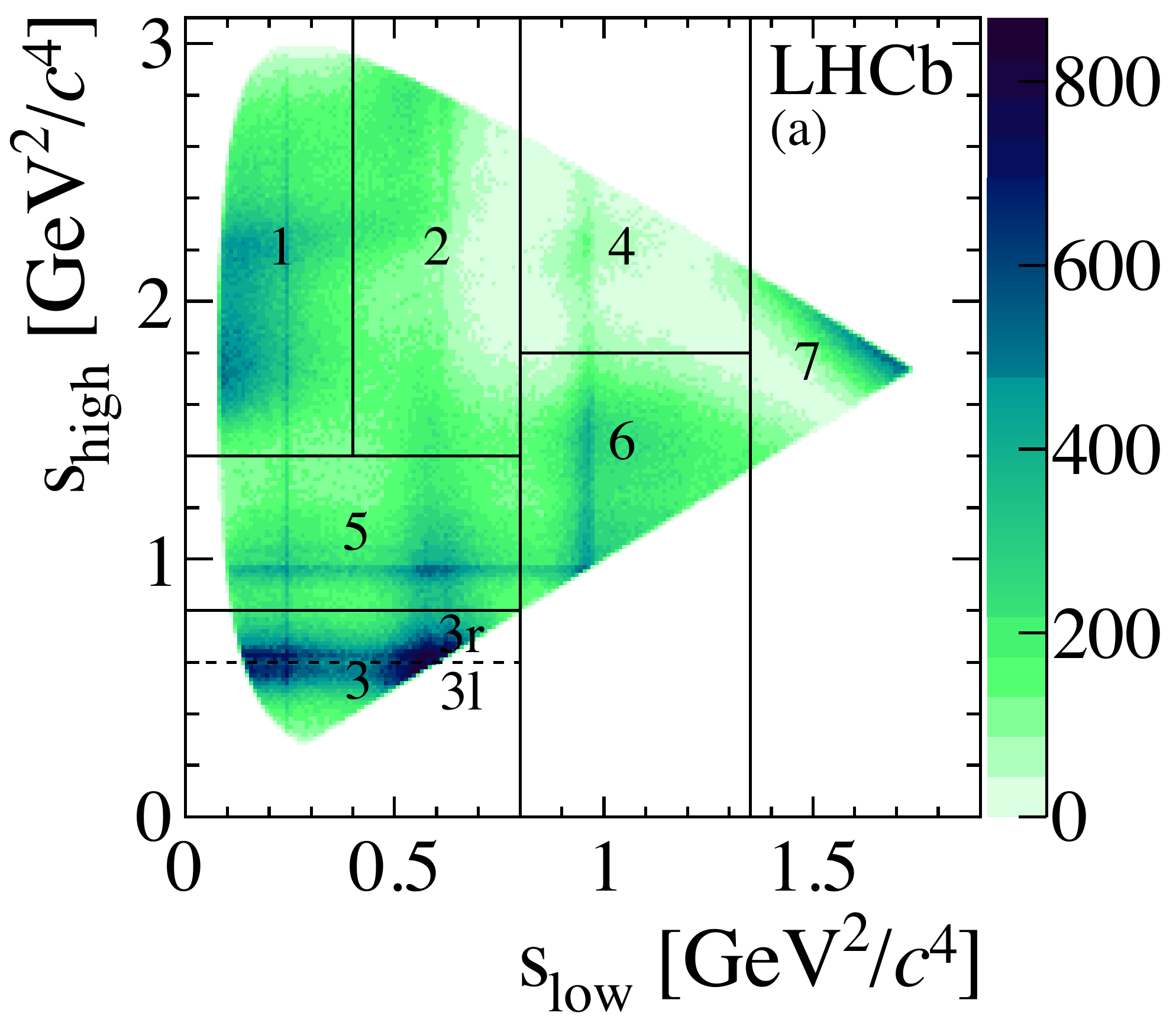}~~~
\includegraphics*[width=0.45\textwidth]{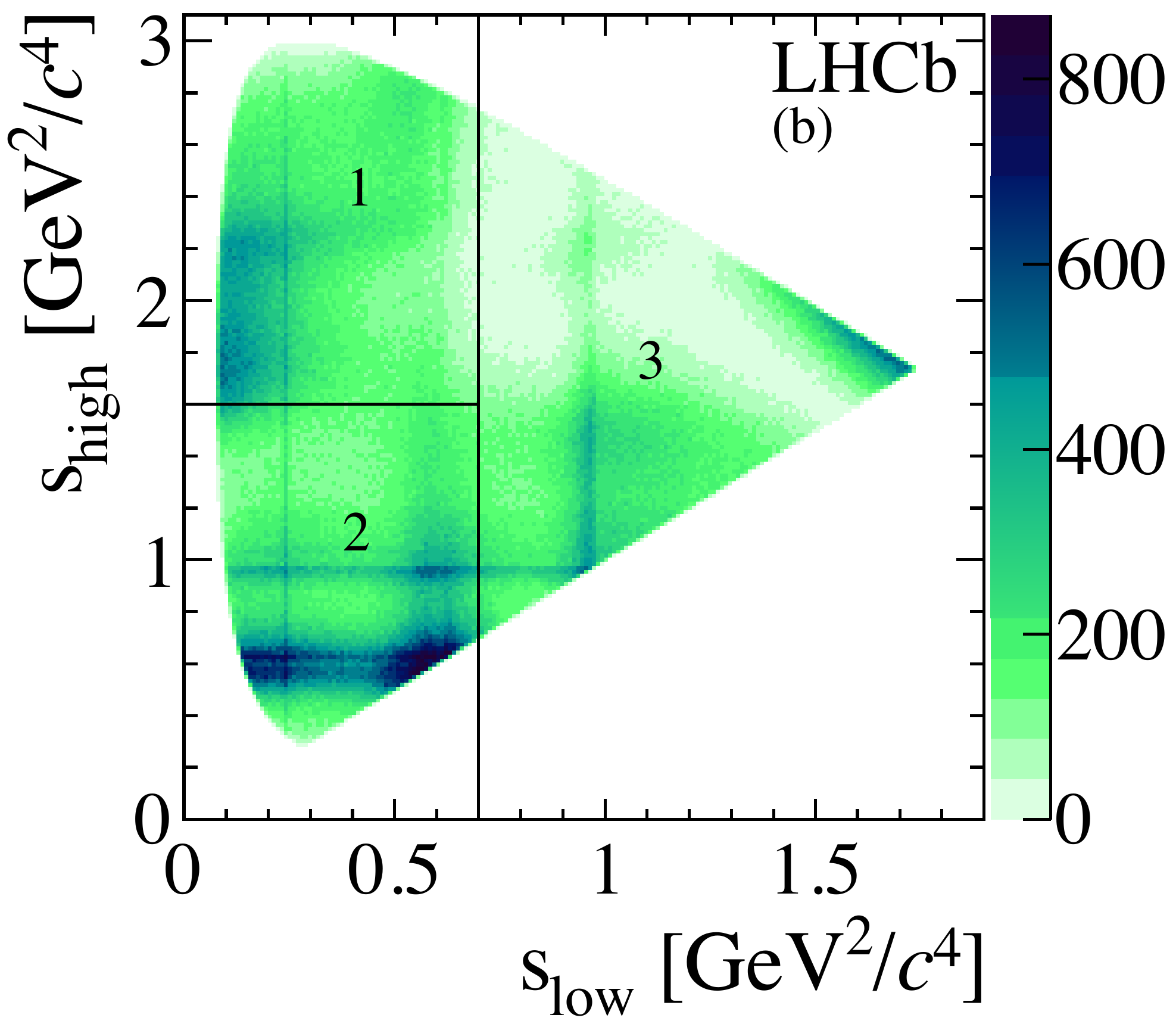} \\
\end{center}
\vspace*{-0.7cm}
\caption{\small Dalitz plot for \Dppp candidates divided into
              (a) seven regions R1-R7 and
              (b) three regions P1-P3. }
  \label{fig:Dregions}
\end{figure}

\begin{figure}[htbp]
\begin{center}
\includegraphics*[width=0.46\textwidth]{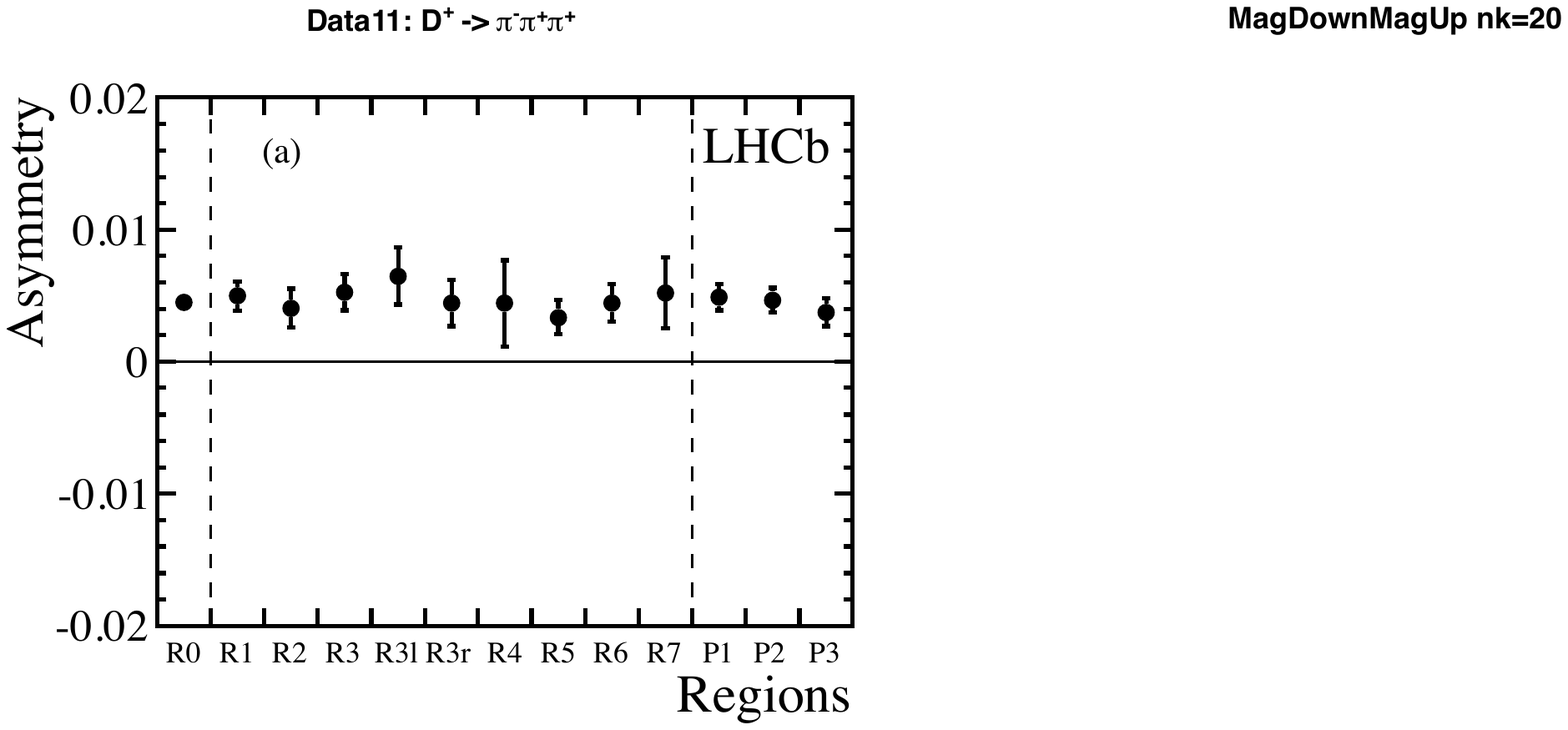}~~~
\includegraphics*[width=0.45\textwidth]{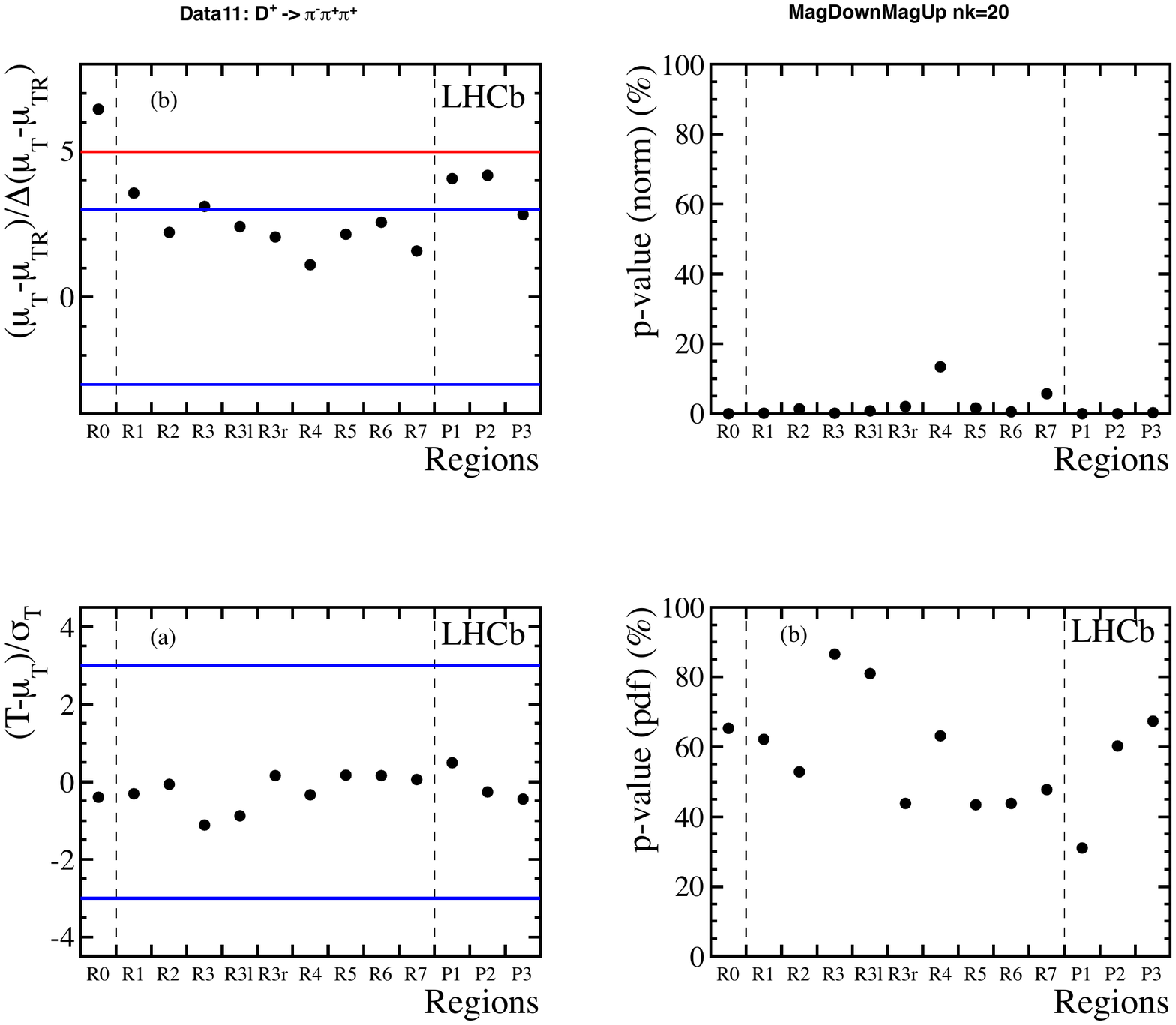}\\
\end{center}
\vspace*{-0.7cm}
\caption{\small (a) Raw asymmetry and (b) the pull values of 
              $\mu_T$ for \Dppp candidates restricted to each region. 
              The horizontal lines in (b)  represent pull values $+3$ and 
              $+5$.
              The region R0 corresponds to the full Dalitz plot.
              Note that the points for the overlapping regions are correlated.}
  \label{fig:DMagDownMagUp_asym}
\end{figure}

The kNN method is applied to the Cabibbo-suppressed 
mode \Dppp with the two region definitions shown in 
Fig.~\ref{fig:Dregions}. To account for the different resonance structure
in $D^+$ and $D^+_s$ decays, the region R1-R7 definition
for the signal mode is different from the definition used in the control mode
(compare Figs.~\ref{fig:Dsregions}a and~\ref{fig:Dregions}a). 
The region P1-P3 definitions are the same.
The results for the raw asymmetry
are shown in Fig.~\ref{fig:DMagDownMagUp_asym}.
The production asymmetry is clearly visible in all 
the regions with the same magnitude as in the control channel
(see Fig.~\ref{fig:DsMagDownMagUp_asym}).
It is accounted for in the kNN method as a deviation of the measured
value of $\mu_T$ from the reference value $\mu_{\it TR}$
shown in Fig.~\ref{fig:DMagDownMagUp_asym}.
In the signal sample the values $\mu_T-0.5=(98\pm 15)\times 10^{-7}$ and
$(\mu_T-\mu_{\it TR})/\Delta(\mu_T-\mu_{\it TR})=6.5\sigma$
in the full Dalitz plot are a consequence of the 0.4\%
global asymmetry similar to that observed in the control mode
and consistent with the previous measurement from LHCb~\cite{LHCb-PAPER-2012-026}.

\begin{figure}[htbp]
\begin{center}
\includegraphics*[width=0.45\textwidth]{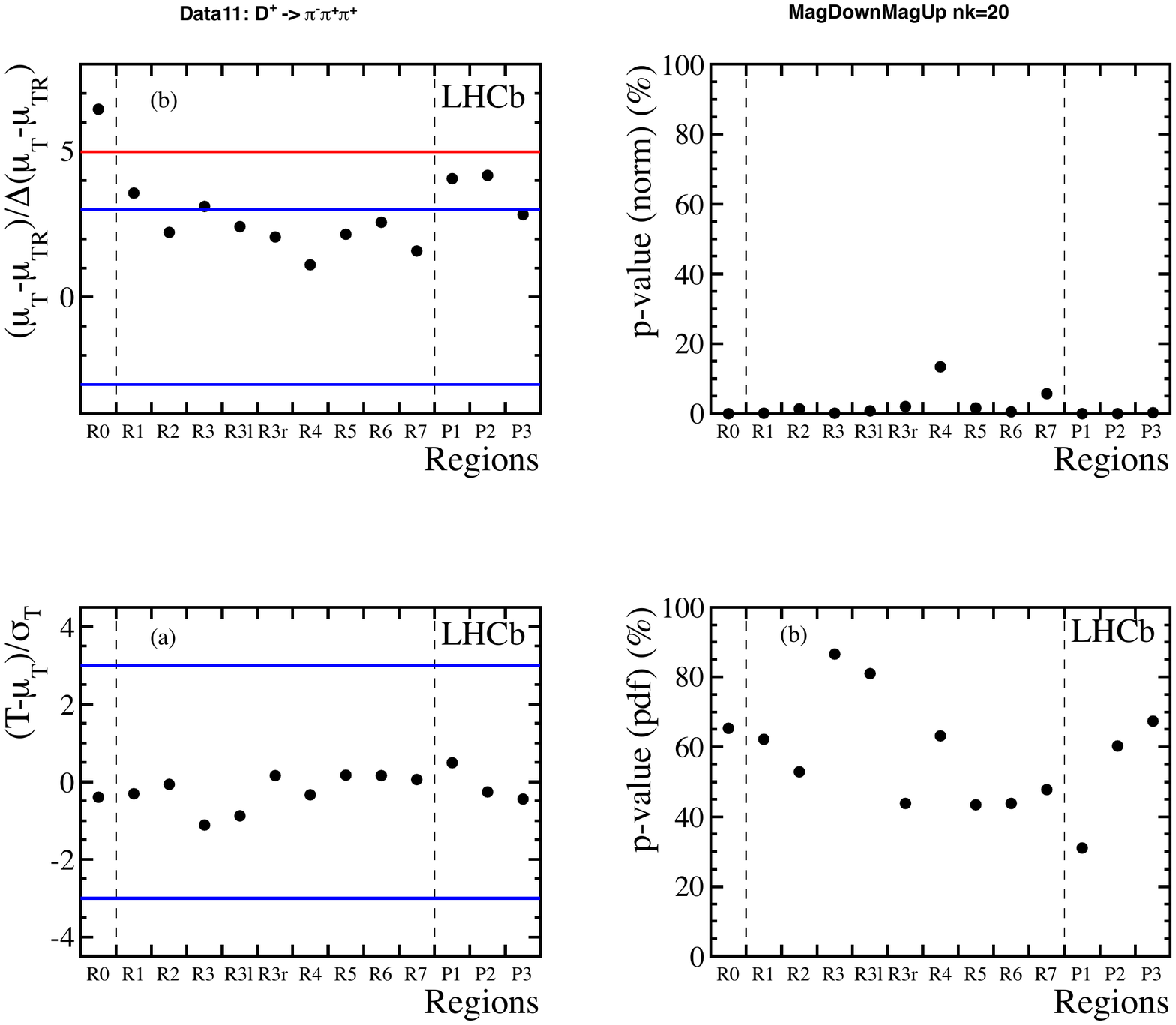}~~~
\includegraphics*[width=0.46\textwidth]{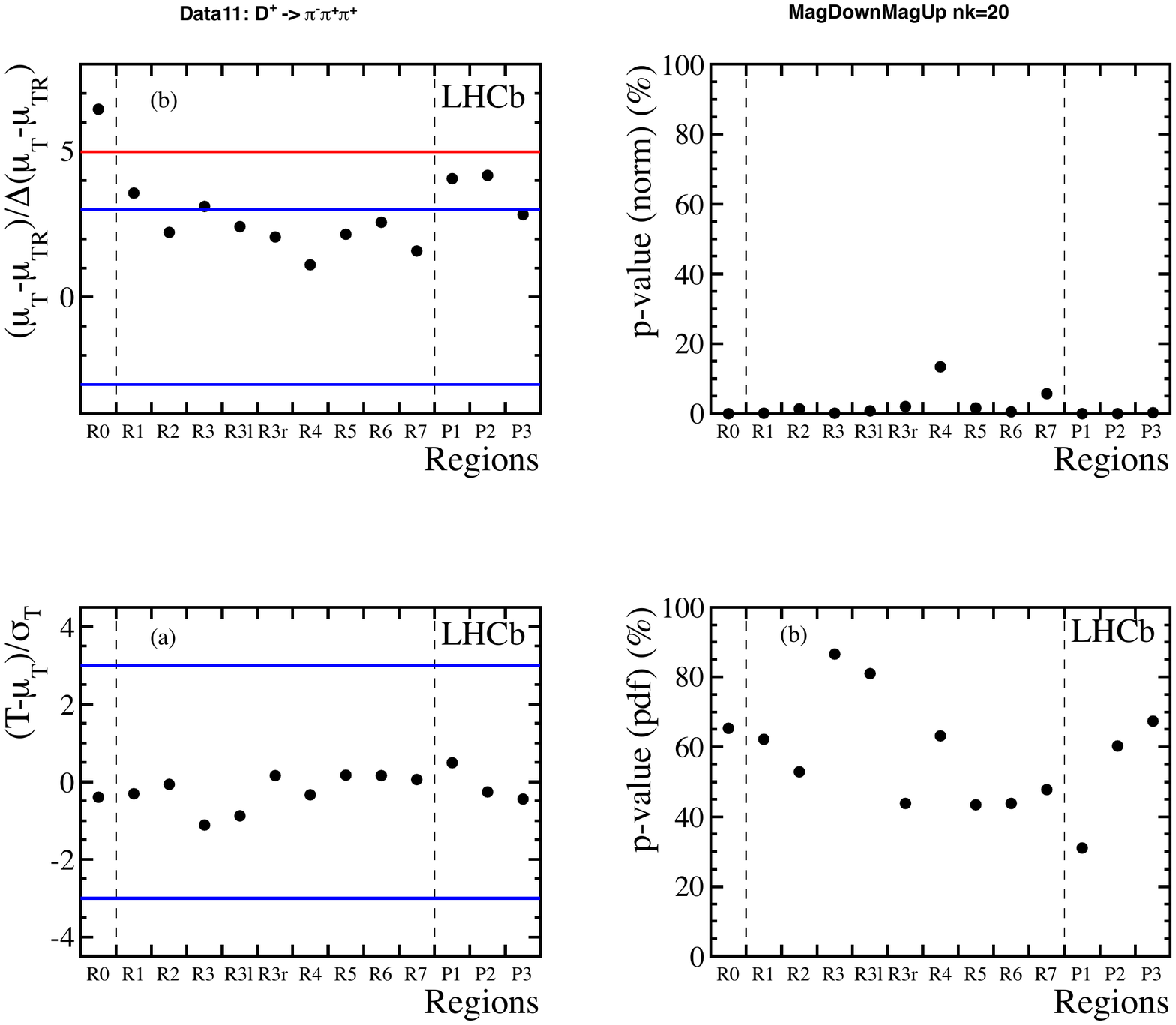} \\
\end{center}
\vspace*{-0.7cm}
\caption{\small (a) Pull values of $T$ and (b) the corresponding p-values
              for \Dppp candidates restricted to each region obtained using the kNN method with $n_k=20$. 
              The horizontal blue lines in (a) represent pull values
              $-3$ and $+3$.  The region R0 corresponds to the full Dalitz plot.
              Note that the points for the overlapping regions are correlated.}
  \label{fig:DMagDownMagUp_nk20}
\end{figure}

The pull values of $T$ and the corresponding p-values for the
hypothesis of no \CPV are shown in Fig.~\ref{fig:DMagDownMagUp_nk20}
for the same regions. To check for any systematic effects, the test is repeated
for samples separated according to magnet polarity. 
Since the sensitivity of the method increases
with $n_k$, the analysis is repeated with $n_k=500$ for all the regions. 
All p-values are above 20\%, 
 consistent with no \CP asymmetry in the signal mode.

\section{Conclusion}
\label{sec:discussion}

A search for \CPV in the decay \Dppp is  performed using 
$pp$ collision data  corresponding to an integrated luminosity of 1.0 fb$^{-1}$
collected by the LHCb experiment at a centre-of-mass energy of 7~TeV. 
Two model-independent methods are applied to a sample 
of 3.1~million \Dppp decay candidates with 82\% signal purity. 

The binned method is based on the study of the local significances \SCPi 
in bins of the Dalitz plot, while the unbinned method uses 
the concept of nearest neighbour events in the pooled \Dp and \Dm sample.
Both methods are also applied to the Cabibbo-favoured \Dsppp decay 
and to the mass sidebands to control possible asymmetries
not originating from \CPV.

No single bin in any of the binning schemes presents an absolute \SCPi  value larger than 3.
Assuming no \CPV, the probabilities of observing local asymmetries 
across the phase-space of the \Dp meson decay as large or larger than 
those in data are above 50\% in all the tested binned schemes. 
In the unbinned method, the p-values are above 30\%. 
All results are consistent with no \CPV.

\section*{Acknowledgements}

\noindent We express our gratitude to our colleagues in the CERN
accelerator departments for the excellent performance of the LHC. We
thank the technical and administrative staff at the LHCb
institutes. We acknowledge support from CERN and from the national
agencies: CAPES, CNPq, FAPERJ and FINEP (Brazil); NSFC (China);
CNRS/IN2P3 and Region Auvergne (France); BMBF, DFG, HGF and MPG
(Germany); SFI (Ireland); INFN (Italy); FOM and NWO (The Netherlands);
SCSR (Poland); MEN/IFA (Romania); MinES, Rosatom, RFBR and NRC
``Kurchatov Institute'' (Russia); MinECo, XuntaGal and GENCAT (Spain);
SNSF and SER (Switzerland); NAS Ukraine (Ukraine); STFC (United
Kingdom); NSF (USA). We also acknowledge the support received from the
ERC under FP7. The Tier1 computing centres are supported by IN2P3
(France), KIT and BMBF (Germany), INFN (Italy), NWO and SURF (The
Netherlands), PIC (Spain), GridPP (United Kingdom). We are thankful
for the computing resources put at our disposal by Yandex LLC
(Russia), as well as to the communities behind the multiple open
source software packages that we depend on.

\newpage
\addcontentsline{toc}{section}{References}
\setboolean{inbibliography}{true}
\bibliographystyle{LHCb}
\bibliography{main,LHCb-PAPER,LHCb-DP}

\end{document}